\documentstyle[prb,aps,epsf,multicol,amssymb]{revtex}

\begin{document}
\draft
\title{
$XY$ models with disorder and symmetry-breaking fields in two
dimensions}

\author{Stefan Scheidl$^{1,2}$ and Michael Lehnen$^{1,3}$}

\address{
$^1$Universit\"at zu K\"oln, Institut f\"ur Theoretische Physik,
Z\"ulpicher Strasse 77, D-50937 K\"oln, Germany
\\
$^2$Materials Science Division, Argonne National Laboratory, Argonne,
Illinois 60439
\\
$^3$Institut f\"ur Plasmaphysik, Forschungszentrum J\"ulich GmbH,
Euratom-Association, D-52425 J\"ulich, Germany
}

\date{February 10, 1998}

\maketitle

\begin{abstract}
 The combined effect of disorder and symmetry-breaking fields on the
 two-dimensional $XY$ model is examined.  The study includes disorder
 in the interaction among spins in the form of random phase shifts as
 well as disorder in the local orientation of the field.  The phase
 diagrams are determined and the properties of the various phases and
 phase transitions are calculated.  We use a renormalization group
 approach in the Coulomb gas representation of the model.  Our results
 differ from those obtained for special cases in previous works.  In
 particular, we find a changed topology of the phase diagram that is
 composed of phases with long-range order, quasi-long-range order, and
 short-range order.  The discrepancies can be ascribed to a breakdown
 of the fugacity expansion in the Coulomb gas representation.
 Implications for physical systems such as planar Josephson junctions
 and the faceting of crystal surfaces are discussed.
\end{abstract}


\begin{multicols}{2}

\section{Introduction}

 We reconsider the two-dimensional $XY$ model with random phase shifts
 and a symmetry-breaking field, which is described by the reduced
 Hamiltonian
\begin{equation}
\label{H} 
{\cal H}=-K_0 \sum_{{\bf r},\alpha} 
\cos (\nabla_\alpha \theta_{\bf r} -A_{{\bf r},\alpha}) 
- H_0 \sum_{\bf r} \cos(p \theta_{\bf r}-\phi_{\bf r}) .
\end{equation} 
 $XY$ spins $\theta_{\bf r}$ are placed on the sites $\bf r$ of a
 square lattice. The coupling of nearest neighbors with a spin
 stiffness $K_0\equiv J/T$ favors a relative angle prescribed by the
 quenched random phase shifts $A$, which we assume to be Gaussian
 distributed with zero mean and variance $\overline {A_{{\bf
 r},\alpha}^2}=\sigma$. The global rotation symmetry of spins is
 broken by the field $H_0$ that selects $p \in {\mathbb N}$ equivalent
 favorable directions for the spins. The model will be considered with
 uniform fields ($\phi=0$) and with random local orientations
 $\phi$ of the field.

 This model describes a variety of physical systems, ranging from
 disordered magnets\cite{RSN83} to Josephson-junction arrays\cite{GK}
 and planar Josephson-junctions.\cite{HG97} It is also closely related
 to surfaces of crystals with bulk disorder\cite{TV90} and even to
 periodic elastic media such as vortex lattices in the presence of a
 pinning potential,\cite{NLS91,HF94} two-dimensional crystals with
 quenched impurities\cite{drN83} or adsorbates on disordered
 substrates.\cite{CN96,CD97a}

 The study of the above and related models started already decades
 ago. We briefly recall some marked steps in its history to provide
 the background for the present work. The $XY$ model in the absence of
 disorder and magnetic fields exhibits the Kosterlitz-Thouless (KT)
 transition from a phase with quasi-long-range order (QLRO, i.e.,
 algebraic decay of the spin-spin correlation) to short-range order
 (SRO, i.e., exponential decay of the spin-spin correlation) at
 $K_\infty = 2/\pi$,\cite{KT73,jmK74,Ber} where the effective spin
 stiffness $K_\infty \lesssim K_0$ is renormalized by thermal
 fluctuations. The transition was identified with an unbinding of
 vortex pairs. Weak {\em uniform} fields have been included by Jos\'e,
 Kadanoff, Kirkpatrick, and Nelson.\cite{JKKN77,PU73} For $p>4$ they
 found that the phase with QLRO is stable against such fields over a
 temperature range $8 \pi/p^2 < K_\infty^{-1} < \pi/2$ and that weak
 fields become relevant at low temperatures $K_\infty^{-1} < 8\pi/p^2
 $ where they stabilize long-range order (LRO, i.e. finite
 magnetization). For $p < 4$ the transition resulting from the
 competition between the ordering tendency of the magnetic field and
 the disordering tendency of the thermal fluctuations could not be
 captured by a KT-like approach.

 Houghton, Kenway, and Ying\cite{HKY81} found the phase with QLRO to
 be unstable to magnetic fields with {\em random} orientation only for
 $K_\infty^{-1} < 4\pi/p^2$.  Accordingly, a phase with QLRO can be
 stable within a finite temperature range $4 \pi/p^2 < K_\infty^{-1} <
 \pi/2$ provided $p > 2 \sqrt 2$.  Their study was extended by Cardy
 and Ostlund\cite{CO82} (CO) to address the nature of the phase
 transitions and of the low-temperature phase. Their analysis implied
 that randomness of the direction of the magnetic field effectively
 generates a new type of disorder on large scales (in the sense of a
 renormalization group), namely random phase shifts
 $A$.\cite{note.bonds}

 Rubinstein, Shraiman, and Nelson\cite{RSN83} (RSN) examined the $XY$
 model with random phase shifts in the absence of fields. They found
 QLRO to be stable against disorder below a certain critical strength
 and a phase diagram with reentrant shape.  Paczuski and
 Kardar\cite{PK91} (PK) added uniform fields and obtained a phase
 diagram composed of various phases with LRO, QLRO, and SRO. The
 transitions between these phases showed multiple reentrance.

 Recent advance in the understanding of the bond-disordered model
 without fields has led to a revised nonreentrant phase diagram, where
 the phase with QLRO turned out to be more stable than found by
 RSN.\cite{NSKL95,CF95,KN96,S97,lhT96} The earlier underestimation of
 the order in the system was caused by a subtle {\em breakdown of the
 fugacity expansion} in the Coulomb gas representation of the model at
 low temperatures.\cite{seK93,MW97} This breakdown was overcome in
 these recent approaches by nonperturbative techniques. The features
 of the revised phase diagram have been confirmed by numerical
 simulations.\cite{KS97,GM97}

 The breakdown of the fugacity expansion is likely to affect {\em all}
 problems involving random phase shifts. In particular one has to ask
 how the structure of the phase diagram, the nature of the phases and
 the nature of the phase transitions are modified. It is also
 important to examine whether a similar breakdown occurs in a
 perturbative treatment of the fields. The purpose of the present work
 is to reexamine model (\ref{H}) in view of these questions.

\begin{figure}
\epsfxsize=0.9 \linewidth
\epsfbox{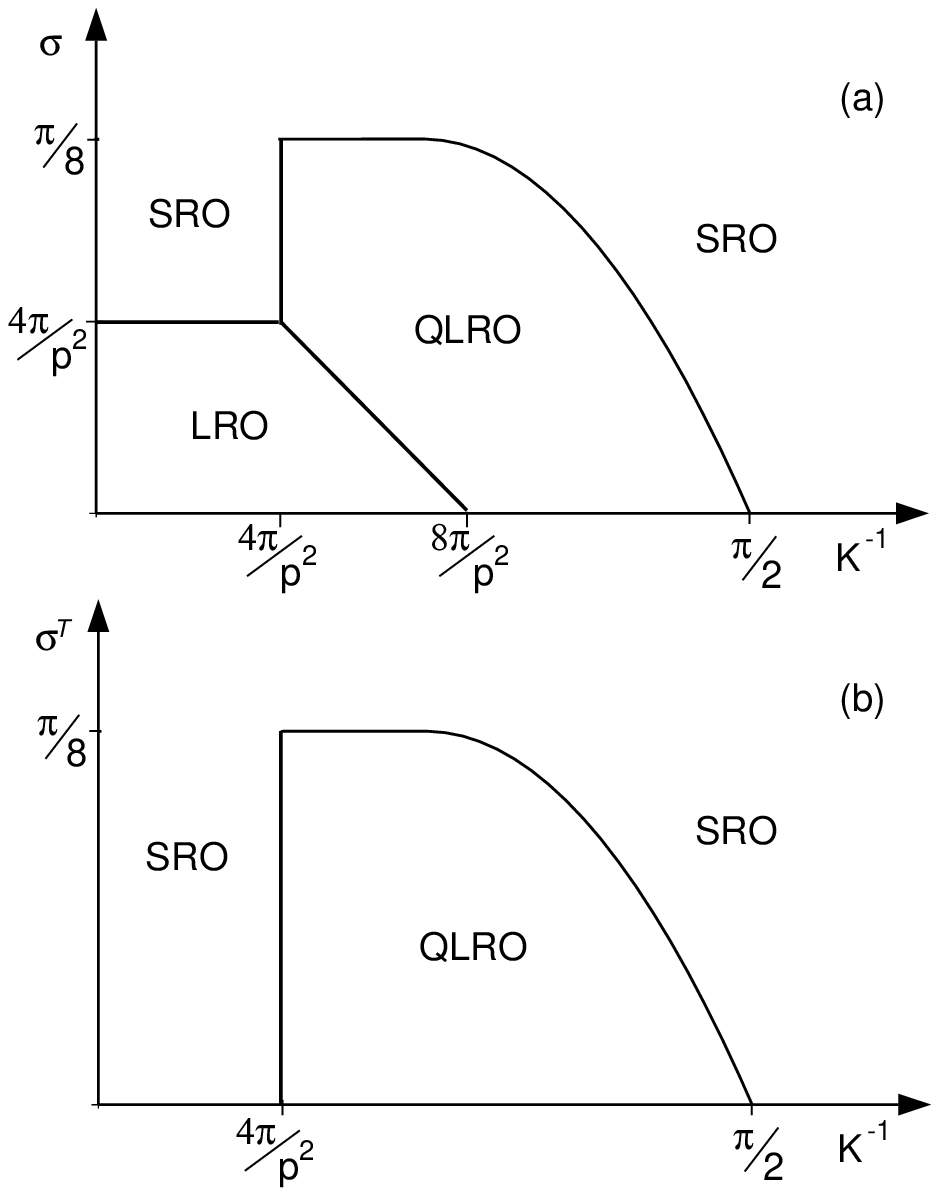}
\narrowtext
\caption{
 Generic phase diagrams (schematically) for the special cases of (a)
 {\em uniform} fields for $p>4$ and (b) {\em random} fields for $p>2
 \sqrt 2$.}
\label{fig.schem}
\end{figure}

 The method of our choice is a mapping of model (\ref{H}) on two
 coupled Coulomb gases, ``vortices'' and ``charges''
 (Sec. II). Thereby we take a pre-averaging over disorder using the
 replica trick. We establish renormalization group flow equations for
 the model parameters to capture the large scale properties of the
 system (Sec. III). From the evaluation of these flow equations
 (Sec. IV) we obtain results that are significantly different from
 those obtained by CO and PK for the special cases of ferromagnetic
 bonds with random fields and bond disorder with uniform fields. Our
 phase diagrams are illustrated in Fig. \ref{fig.schem} for these two
 previously examined special cases.  We conclude with a discussion of
 our results and of consequences for physical realizations of the $XY$
 model (Sec. V). Appendices with some calculational details and a
 dictionary for a translation of notations from previous works to the
 present work are included.

\section{Model transformations}

 In this Section we specify the stochastic properties of the random
 phase shifts considered subsequently. In addition, we transfer
 randomness in the orientation of the symmetry-breaking fields to the
 random phase shifts such that we need to deal only with (correlated)
 bond disorder. The latter model is then mapped onto two coupled
 Coulomb gases. Eventually we arrive at a replica representation of
 the model that serves as basis for our further analysis.

\subsection{Disorder specification}

 In order to be able to treat the model irrespectively of a randomness
 in the local field orientation, we decompose the field ${\bf A}\equiv
 \{A_\alpha\}$ in Fourier space into its longitudinal and transverse
 component,
\begin{mathletters} 
\label{def.A}
\begin{eqnarray}
{\bf A} &=&{\bf A}^L + {\bf A}^T,
\\
A^L_{{\bf k};\alpha} &=& 
\sum_\beta P^L_{{\bf k};\alpha,\beta} A_{{\bf k};\beta} \equiv
\sum_\beta \frac  {k_\alpha k_\beta} {k^2} A_{{\bf k};\beta},
\\
A^T_{{\bf k};\alpha} &=& 
\sum_\beta P^T_{{\bf k};\alpha,\beta} A_{{\bf k};\beta} \equiv
\sum_\beta \left(1- \frac{k_\alpha k_\beta}{k^2} \right) A_{{\bf k};\beta}.
\end{eqnarray} 
\end{mathletters} 
 Both components (polarizations $\Pi=L,T$) are supposed to be Gaussian
 distributed with zero average and to have variances (over-lining
 denotes averaging over quenched disorder)
\begin{equation} 
\label{corr.A} 
\overline{ A^\Pi_{{\bf k};\alpha} A^\Pi_{{\bf k}';\beta}} =
\sigma^\Pi P^\Pi_{{\bf k};\alpha,\beta} \delta({\bf k}+{\bf k}') ,
\end{equation}
 where we explicitly allow for $\sigma^L \neq \sigma^T$. Then the
 random phase shifts have a variance $\sigma = ( \sigma^L +
 \sigma^T)/2$ and are in general correlated in space. Only in the
 special case $\sigma^L = \sigma^T$ the phase shifts are uncorrelated
 on different bonds of the lattice.

 Neglecting the effect of vortices and fields completely, the
 spin-spin correlation (angular brackets denote a thermal average for
 a fixed disorder realization)
\begin{equation}
\Gamma_p({\bf r}) \equiv  
\overline{\langle \cos[ p\theta_{\bf r} - p \theta_{\bf 0}] \rangle}
\label{def.Gamma}
\end{equation}
 can be calculated in the spin-wave approximation and is found to
 decay algebraically,
\begin{equation}
\Gamma_p({\bf r}) \sim r^{-\eta},
\label{Gamma.sw}
\end{equation}
 with an exponent $\eta=(p^2/2\pi)(1/K +\sigma^L)$.  In this
 approximation $\eta$ depends only on the {\em longitudinal} disorder.
 As in Ref. \CITE{CO82} we will also examine correlations of the
 Edwards-Anderson order parameter
\begin{mathletters}
\label{EA.corr}
\begin{eqnarray}
\psi^{ab}({\bf r}) &\equiv& e^{i\theta^a_{\bf r} - i\theta^b_{\bf r}}
\\
\overline {\langle {\psi^{ab}}^*({\bf r}) \psi^{ab}({\bf 0})  \rangle}
&\sim& r^{-\overline{\eta}}
\end{eqnarray}
\end{mathletters}
 between different replicas $a \neq b$ of the system. In the spin-wave
 approximation the exponent $\overline \eta=1/\pi K$ is completely
 independent of disorder.\cite{CO82}

\subsection{Random fields}
\label{sec.map}

 A convenient way of dealing with the randomness in the local field
 orientation $\phi$ is to eliminate the angles $\phi$ in favor of
 modified phase shifts $\tilde {\bf A}={\bf A}- {\bbox \nabla}
 \phi/p$ by a substitution $\theta = \tilde \theta +\phi/p$ . This
 transformation leaves the thermodynamic partition sum invariant,
 i.e.,
\begin{equation} 
\label{Z}
Z(\{{\bf A} \},\{\phi \}) \equiv \int {\cal D}\theta \  e^{-{\cal H}} =
Z(\{\tilde{\bf A}\},\{0\}).
\end{equation}
 Therefore, the existence and location of phase transitions in the
 phase diagram are not affected by this manipulation. In contrast,
 spin correlation functions are {\em not} invariant and can display
 qualitatively different large-scale behavior before and after the
 transformation. 

 In order to restrict our analysis of random phase shifts to those
 with correlations (\ref{corr.A}) even after the transformation, we
 limit the possible distributions of the angles $\phi$ to those
 having a Gaussian distribution with correlations
\begin{equation}
\label{corr.alpha}
\overline{\phi_{\bf k} \phi_{{\bf k}'}}= 
\Delta \frac{1}{k^2} \delta({\bf k} + {\bf k}').
\end{equation}
 Only such random orientations are transformed by the above mapping
 into longitudinal phase shifts with a variance
\begin{equation}
\label{sig.til}
\tilde \sigma^L = \sigma^L+\frac{\Delta}{p^2}.
\end{equation}
 Equation (\ref{corr.alpha}) implies correlations
 $\overline{(\phi_{\bf r} - \phi_{\bf 0})^2} \approx (\Delta/\pi) \ln
 r$ for $r \gg 1$ in real space.  However, in the limit $\Delta \to
 \infty$ the relative fluctuations even of neighboring angles are that
 large that they can be considered practically as uncorrelated since
 they enter the Hamiltonian only modulo $2 \pi$. In this sense the
 examination of the model (\ref{H}) with random phase shifts and
 uniform field effectively includes the model with random
 symmetry-breaking fields in the limit $\sigma^L \to \infty$.

\subsection{Coulomb gas representation}
 
 We apply standard techniques\cite{JKKN77} to map model (\ref{H}) onto
 a Coulomb gas. For this purpose we first use the Villain
 approximation\cite{Vil75} to replace both cosines in Eq. (\ref{H}) by
 a slightly different periodic potential and obtain an effective
 Hamiltonian
\begin{eqnarray} 
\label{H.vil} 
{\cal H}&=&\frac {K}2 \sum_{\bf r} 
({\bbox \nabla} \theta_{\bf r}- {\bf A}_{\bf r}+2 \pi {\bf M}_{\bf r})^2 
\nonumber \\
&&+ \frac{H}2 \sum_{\bf r} (p \theta_{\bf r} +2\pi m_{\bf r})^2.
\end{eqnarray}
 The partition sum of this effective model contains not only the
 integration over the angles $\theta$ but also a summation over the
 integer-valued vector field ${\bf M}$ and the integer-valued scalar
 field $m$. The parameters of this effective model are related to the
 original parameters by $K =f(K_0)$ and $H =f(H_0)$ through the
 function \cite{Vil75,JKKN77}
\begin{equation} 
\label{def.f}
f(x) \approx \left \{
\begin{array}{ll}
  |x| & \quad {\rm for} \quad  x \gg 1, \\ 
  -1/\ln(x^2/4) & \quad {\rm for} \quad  x \ll 1.
\end{array} \right.
\end{equation}
 Since we are mainly interested in the stability of the phase with
 QLRO (i.e., $K_0 \gtrsim 1$) to symmetry-breaking fields (i.e., $H_0
 \ll 1$), we effectively operate in the limiting cases $K \approx K_0$
 and $H \approx -1/ \ln(H_0^2/4)$.

 For small $H$ the sum over $m$ converges very slowly. Therefore it is
 of advantage to go over from the field $m$ to a conjugate
 integer-valued field $n$ by means of the Poisson summation formula
 (some calculational details are given in Appendix \ref{app.cg}).
 After integrating out the angles $\theta$ in the partition sum we end
 up with a new effective Hamiltonian for two coupled Coulomb
 gases. The first Coulomb gas is the gas of {\em vortices} with a
 vorticity $N= {\bbox \nabla} \times {\bf M} \in {\mathbb Z} $. These
 vortices are located on the dual lattice, i.e., on plaquette centers
 ${\bf R}$.\cite{JKKN77} The {\em ``charges''} $n_{\bf r} \in {\mathbb
 Z}$ constitute the second Coulomb gas. These Coulomb gases are
 coupled to {\em ``quenched vortices''} $Q_{\bf R} \equiv (1/2
 \pi){\bbox \nabla} \times {\bf A}$ and {\em ``quenched charges''}
 $q_{\bf r} \equiv (K/p) {\bbox \nabla} \cdot {\bf A}$, which are
 generated by the transverse and longitudinal components of the bond
 disorder respectively.

 The Hamiltonian of the effective model reads
\begin{mathletters} 
\label{H.rs}
\begin{eqnarray} 
{\cal H}&=&{\cal H}_{\rm v} + {\cal H}_{\rm c} + {\cal H}_{\rm vc} 
+ {\cal H}_{\rm x},
\\
{\cal H}_{\rm v} &=& - \frac p2 \sum_{{\bf R} \neq {\bf R}'} \frac {2 \pi K }p 
\ln |{\bf R} - {\bf R}'| (N_{\bf R}-Q_{\bf R}) (N_{\bf R'}-Q_{\bf R'}) 
\nonumber \\
&& +E_{\rm v} \sum_{\bf R} N_{\bf R}^2  ,
\\
{\cal H}_{\rm c} &=&- \frac p2 \sum_{{\bf r} \neq {\bf r}'} \frac p {2 \pi K } 
\ln |{\bf r} - {\bf r}'| (n_{\bf r}-i q_{\bf r}) (n_{\bf r'}-i q_{\bf r'})  
\nonumber \\
&& +E_{\rm c} \sum_{\bf r} n_{\bf r}^2 ,
\\
{\cal H}_{\rm vc} &=& i p  \sum_{\bf R,r} \omega({\bf R}-{\bf r}) 
N_{\bf R} n_{\bf r},
\label{H.vc} \\
{\cal H}_{\rm x} &=& -\frac p2 \sum_{{\bf r} \neq {\bf r}'} \frac p {2 \pi K } 
\ln |{\bf r} - {\bf r}'| q_{\bf r} q_{\bf r'} .
\label{H.x}
\end{eqnarray}
\end{mathletters}
 It has been split into the interaction among vortices $ {\cal H}_{\rm
 v}$, the interaction among charges $ {\cal H}_{\rm c}$, the
 interaction between vortices and charges $ {\cal H}_{\rm vc}$, and an
 excess term ${\cal H}_{\rm x}$.  Particles of the same type interact
 logarithmically and the coupling between the two gases is given by
 the angle 
\begin{equation}
\omega({\bf R}-{\bf r}) = \arctan ({\bf R}-{\bf r})
\end{equation}
 enclosed by their relative distance and an arbitrary reference
 direction in the $xy$ plane (see Appendix \ref{app.cg} for some
 intermediate steps).  Due to the introduction of the charges by means
 of the Poisson summation formula there are two imaginary-valued
 contributions to the Hamiltonian: the coupling of thermal charges $n$
 to quenched charges $q$ and the coupling of vortices $N$ to charges
 $n$. However, the partition sum is real-valued since the imaginary
 parts cancel in pairs $\exp[-{\cal H} (\{N\}, \{n\})] + \exp[-{\cal
 H} (\{N\}, \{-n\})]$.

 Both gases are dilute for $K \gg1$ and $H \ll 1$, where the core
 energies
\begin{equation} 
\label{defs.E}
E_{\rm v} \equiv \pi \gamma K, \ \ \ 
E_{\rm c} \equiv \frac 1{2 H } + \frac{p^2 \gamma}{4 \pi K } 
\end{equation}
 are large ($\gamma \approx 1.6$ is a constant of order
 unity\cite{note.gamma} emerging from the lattice Green
 function). Although these core energies are determined by $K$ and $H$
 we will consider them formally as independent parameters in the
 subsequent treatment.

 The partition function of this effective model (\ref{H.rs}) reads
\begin{equation} 
\label{Z.rs}
Z \equiv Z(2 \pi K/p, E_{\rm v}, E_{\rm c},\{Q\},\{q\})=
\sum_{\{N\},\{n\}}' e^{-{\cal H}}.
\end{equation}
 The primed summation is constrained by neutrality conditions
\begin{equation}
\label{neutr}
\sum_{\bf R} N_{\bf R}=0, \ \ \ 
\sum_{\bf r} n_{\bf r}=0,
\end{equation}
 since the real part of the energy of nonneutral configurations
 diverges logarithmically with the system size.

\subsection{Replication}

 The direct analysis of the partition sum (\ref{Z.rs}) in the spirit
 of the KT renormalization group (RG) is prevented by the broken
 translation invariance in the presence of the quenched vortices and
 charges. Since we are aiming at the calculation of disorder-averaged
 quantities, such as the correlation function (\ref{def.Gamma}), we
 may perform the disorder average on the replicated partition sum
 already at this stage.\cite{EA75}

 For an integer number of replicas $n$ the Hamiltonian of the
 replicated system is most conveniently expressed in a vector
 notation, where ${\bf n}_{\bf r} \equiv \{n^a_{\bf r}\}$ for
 $a=1,\dots,n$ is a replica charge at site ${\bf r}$ and ${\bf N}_{\bf
 R} \equiv \{N^a_{\bf R}\}$ is a replica vortex at site ${\bf R}$:
\begin{mathletters} 
\label{H.rep} 
\begin{eqnarray} 
{\cal H}&=&{\cal H}_{\rm v}+{\cal H}_{\rm c}+{\cal H}_{\rm vc},
\\
{\cal H}_{\rm v} &=& - \frac 12 \sum_{{\bf R}\neq {\bf R}'}
{\bf N}_{\bf R} \cdot {\bf K}_{\rm v} \cdot {\bf N}_{\bf R'} 
2 \pi \ln |{\bf R}-{\bf R'}| 
\nonumber \\
&&+\sum_{\bf R} {\bf N}_{\bf R} \cdot {\bf E}_{\rm v} \cdot {\bf
N}_{\bf R} 
\\
{\cal H}_{\rm c} &=& - \frac 12 \sum_{{\bf r}\neq {\bf r}'} 
{\bf n}_{\bf r} \cdot {\bf K}_{\rm c} \cdot {\bf n}_{\bf r'} 
2 \pi \ln |{\bf r}-{\bf r'}|
\label{H.rep.Hc} \nonumber \\
&&+\sum_{\bf r} {\bf n}_{\bf r} \cdot {\bf E}_{\rm c} \cdot {\bf
n}_{\bf r} ,
\\
{\cal H}_{\rm vc} &=& i p  \sum_{\bf R,r} 
\omega({\bf R}-{\bf r}) {\bf N}_{\bf R} \cdot  {\bf n}_{\bf r}.
\end{eqnarray}
\end{mathletters} 
 All parameters are incorporated in the replica-symmetric coupling
 matrices
\begin{mathletters} 
\label{E.K} 
\begin{eqnarray} 
K_{\rm v}^{ab}&=&K \delta^{ab}- \frac{\sigma^T K^2}{1+n \sigma^T K},
\\
K_{\rm c}^{ab}&=&\frac{p^2}{4 \pi^2} 
\left(\frac 1{K } \delta^{ab} + \sigma^L\right) ,
\\
E_{\rm v}^{ab}&=&\pi \gamma K_{\rm v}^{ab},
\\
E_{\rm c}^{ab}&=&\frac 1{2H} \delta^{ab} + \pi \gamma K_{\rm c}^{ab}.
\end{eqnarray}
\end{mathletters} 

 Hamiltonian (\ref{H.rep}) is a generalization of that given in
 Sec. II of CO. In the limit $\sigma^L \to \infty$ we retrieve their
 model, i.e., Eq. (\ref{H.rep.Hc}) becomes equivalent to their
 Eq. (2.15). Furthermore, in this limit the contribution of the charge
 core energy proportional to $\sigma^L$ enforces an additional
 neutrality 
\begin{equation}
\label{neu.rep}
 \sum_a n_{\bf r}^a=0 
\end{equation}
 across the replicas, i.e., the weight of nonneutral configurations
 vanishes in the partition sum. Then the vector ${\bf n}$ can be
 decomposed into pairs of type $(0,\dots,0,1,0,\dots,-1,0,\dots0)$ at
 every site. Such pairs correspond to the vector charges of
 CO. However, for finite $\sigma^L$ the summation over charges is {\em
 not} restricted by this additional neutrality condition. Note that
 Eq. (\ref{H.rep}) contains an {\em explicit} coupling between
 vortices and charges.

\subsection{Duality}

 The effective model under consideration is
 self-dual,\cite{JKKN77,PK91} i.e., it is invariant under a mutual
 exchange of vortices and charges with a simultaneous substitution of
 parameters.  For the unreplicated model (\ref{H.rs}) these
 substitutions are
\begin{equation}
\label{dual}
{\bf R} \leftrightarrow {\bf r}, \
N \leftrightarrow n, \
Q \leftrightarrow iq,\
\frac {2 \pi K} p \leftrightarrow \frac p {2 \pi K}, \
E_{\rm v} \leftrightarrow E_{\rm c}.
\end{equation}
 This duality implies a relation
\begin{eqnarray} 
\label{dual.Z}
 Z(2 \pi K/p, E_{\rm v}, E_{\rm c},\{Q\},\{q\}) 
e^{+{\cal H}_x(2 \pi K/p,\{q\})}= 
\nonumber \\
 Z(p/2 \pi K, E_{\rm c}, E_{\rm v},\{iq\},\{-iQ\})
e^{+{\cal H}_x(p/2 \pi K,\{-iQ\})}
\end{eqnarray}
 for the partition sum,\cite{PK91} which simplifies our subsequent
 calculations because it enables us to relate renormalization effects
 arising from vortex fluctuations to renormalization effects arising
 from charge fluctuations or vice versa. In this way PK obtained
 recursion relations for the bond-disordered model {\em with} field
 from those of RSN {\em without} field. We are going to use the same
 strategy starting from the corrected flow equations.\cite{S97}

 We will proceed within the replica formulation (\ref{H.rep}), where
 the self-duality of the model is reflected by the substitutions
\begin{equation}
\label{dual.rep}
{\bf R} \leftrightarrow {\bf r}, \
{\bf N} \leftrightarrow {\bf n}, \
{\bf K}_{\rm v} \leftrightarrow {\bf K}_{\rm c}, \ 
{\bf E}_{\rm v} \leftrightarrow {\bf E}_{\rm c}.
\end{equation}
 The excess contribution (\ref{H.x}) to the Hamiltonian of the
 unreplicated system effectively modifies the action of the
 longitudinal disorder distribution.  The duality ${\bf K}_{\rm v}
 \leftrightarrow {\bf K}_{\rm c}$ implies $\sigma^{L} \leftrightarrow
 - (4 \pi^2 K^2/p^2) \sigma^{T}/(1+n \sigma^T K)$. The latter relation
 is consistent with Eq. (\ref{dual}) if one keeps in mind the presence
 of the excess term in Eq. (\ref{dual.Z}). This term gives rise to the
 denominator $1+n \sigma^T K$ that becomes unity in the limit $n \to
 0$ that eventually will be taken.

\section{Renormalization}

 After the Coulomb gas representation of our model has been
 established in the previous Section we now turn to a scaling analysis
 for the relevance of vortices and charges. We basically follow
 Kosterlitz' RG approach \cite{jmK74} to the $XY$ model and its recent
 generalization\cite{S97} to the $XY$ model with random phase
 shifts. To account for magnetic fields we have to include charges and
 their coupling to the vortices.

 Following Ref. \CITE{S97} we introduce the notion of ``{\em types}''
 of replica vortices and charges. We consider charges at site ${\bf
 r}$ and site ${\bf r}'$ as being of the same or of different type if
 ${\bf n}_{\bf r} = {\bf n}_{{\bf r}'}$ or ${\bf n}_{\bf r} \neq {\bf
 n}_{{\bf r}'}$ in analogy to the vortex types. From the core energy
 contribution to the Hamiltonian (\ref{H.rep}) we identify fugacities
\begin{mathletters}
\label{fuga}
\begin{eqnarray}
y_{\rm v}({\bf N}) &=& 
e^{- {\bf N} \cdot {\bf E}_{\rm v} \cdot {\bf N}}
\\
y_{\rm c}({\bf n}) &=& 
e^{- {\bf n} \cdot {\bf E}_{\rm c} \cdot {\bf n}}
\end{eqnarray}
\end{mathletters}
 for each type of particles. Because of the replica symmetry of the
 matrices ${\bf E}_{\rm v}$ and ${\bf E}_{\rm c}$ different particle
 types have the same fugacity if they belong to the same ``{\em
 class},'' that is, if they are of the same type after a permutation
 of replicas and/or an inversion of the replica vector (such as ${\bf
 n}_{\bf r} \to - {\bf n}_{\bf r}$).

\subsection{Scaling analysis}
\label{sec.scal}

 In the limit of vanishing field all charge fugacities (\ref{fuga}b)
 go to zero.  The relevance of a weak field can be examined by looking
 at the scaling behavior of the charge fugacities. As soon as the
 fugacity of some charge type increases under rescaling, the field
 constitutes a relevant perturbation. Analogously, the relevance of
 vortices signals the importance of topologically nontrivial spin
 configurations.  However, while a relevance of vortices indicates a
 tendency to reduced spin correlations on large scales, a relevance of
 charges expresses a tendency to enhanced spin correlations.

 A rescaling of lengths ${\bf R} \to e^{-dl} {\bf R}$ and ${\bf r} \to
 e^{-dl} {\bf r}$ with an infinitesimal increase on a logarithmic
 length scale $l$ has the following effect on Hamiltonian
 (\ref{H.rep}): the angles $\omega$ are invariant but the logarithms
 generate additive constants, $\ln|\dots| \to \ln|\dots| - dl$. Using
 the neutrality conditions (\ref{neutr}) the twofold spatial
 summations over these constants can be reduced to single
 summations. Then the rescaling effectively amounts to an additive
 flow of the core matrices ${\bf E}$, or equivalently to a
 multiplicative flow of the fugacities
\begin{mathletters}
\begin{eqnarray}
\label{flow_yv}
\left. \frac{d}{dl} \right|_{\rm scal} y_{\rm v}({\bf N}) &=& 
(2- \pi {\bf N} \cdot {\bf K}_{\rm v}
   \cdot {\bf N}) y_{\rm v}({\bf N}),  \\
\label{flow_yc}
\left. \frac{d}{dl} \right|_{\rm scal} y_{\rm c}({\bf n}) &=& 
(2- \pi {\bf n} \cdot {\bf K}_{\rm c}
   \cdot {\bf n}) y_{\rm c}({\bf n}) .
\end{eqnarray}
\end{mathletters}
 For a finite number $n$ of replicas we can now determine the
 stability of QLRO to vortices and charges at given temperature
 $K^{-1}=T/J$ and disorder strengths $\sigma^\Pi$ by looking for the
 most relevant class of particles. We focus here on the charge sector,
 since the vortex sector has been covered in Ref. \CITE{S97}. The {\em
 most relevant} charge class is the one that minimizes
\begin{equation}
{\bf n} \cdot {\bf K}_{\rm c} \cdot {\bf n}=
\frac{p^2}{4 \pi^2} \left[\sum_a (n^a)^2 K^{-1} + 
\left( \sum_a n^a \right)^2 \sigma^L \right].
\end{equation}
 The minimum is provided by charges of class $\uparrow$ for $\sigma^L
 < K^{-1}$ and by charges of class $\uparrow \downarrow$ for $\sigma^L
 > K^{-1}$.  The class ``$\uparrow$'' is composed of all replica
 charges ${\bf n}$ that contain an elementary charge in a single
 replica [e.g., ${\bf n}=(0,\dots,0,\pm 1,0,\dots,0)$] and the class
 ``$\uparrow \downarrow$'' is composed of all replica charges ${\bf
 n}$ that contain an elementary charge in  one replica and an
 opposite elementary charge in a second replica [i.e., the CO vector
 charges, such as ${\bf n} = (0,\dots,0,+1,0,\dots,0,-1,0,\dots,0)$].

\begin{figure}
\epsfxsize=0.9 \linewidth
\epsfbox{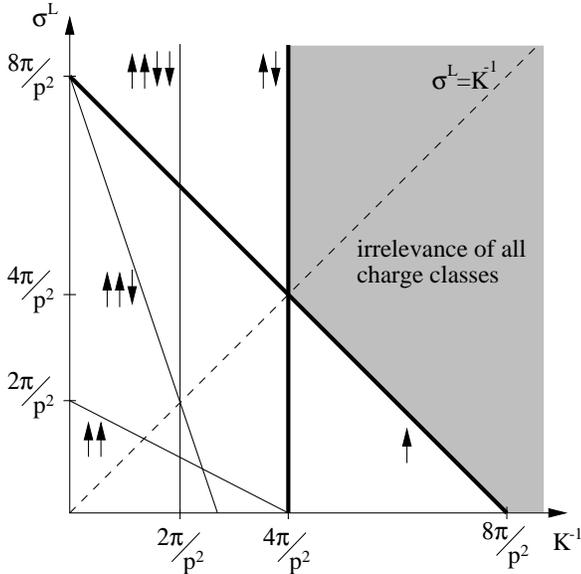}
\narrowtext
\caption{
 This diagram indicates the location of the lines on the
 low-temperature side of which various charges classes become relevant
 according to the scaling (\ref{flow_yc}). In the shaded area {\em
 all} charges are irrelevant. In the whole parameter plane the {\em
 most relevant} charge class is either $\uparrow$ (on the
 high-temperature side of the dashed line at $K^{-1}=\sigma^L$) or
 $\uparrow \downarrow$ (on the low-temperature side of the dashed
 line).}
\label{fig.relev}
\end{figure}

 Let us consider for illustration vector charges that are composed
 only of $s$ elementary charges, thereof $s_\uparrow$ are positive and
 $s_\downarrow$ are negative. The fugacity of such charges scales like
\begin{equation}
\label{mo.re.c}
y_{\rm c}({\bf n}) \sim \exp \left\{l\left[2-
\left[(s_\uparrow + s_\downarrow) K^{-1}
 - (s_\uparrow -s_\downarrow)^2 \sigma^L\right] 
\frac{p^2}{4 \pi}\right]\right\} .
\end{equation}
 Figure \ref{fig.relev} displays the stability boundary in the
 ($K^{-1},\sigma^L$) plane for various classes of charges, which are
 irrelevant on the high-temperature side of the corresponding (full)
 lines. Charge class $\uparrow$ is most relevant on the
 high-temperature side ($\sigma^L < K^{-1}$) and charge class
 $\uparrow \downarrow $ is most relevant on the low temperature side
 ($\sigma^L > K^{-1}$) of the dashed line.

 There is an important difference in the scaling analysis for charges
 and for vortices. The region of the phase diagram, where {\em all}
 $n$-component charges are irrelevant, is {\em independent} of the
 number $n$ of replicas. In addition, the boundary of this region is
 invariantly given for {\em all} integer $n$ by the charge classes
 $\uparrow$ or $\uparrow \downarrow$ (strictly speaking, the class
 $\uparrow \downarrow$ exists only for $n \geq 2$). This independence
 on $n$ indicates that the replica limit $n \to 0$ can be taken in a
 very simple way. In contrast, the region where all vortices are
 irrelevant depends explicitly on $n$ and so does the class of the
 most relevant vortices at lower temperatures (see Fig. 2 of
 Ref. \CITE{S97}). Therefore the replica limit will be easier to take
 for charges than for vortices. The save proceeding for both cases is
 to consider the collective renormalization of the physical parameters
 by all classes of vortices and charges, which will be carried out
 subsequently.

 We anticipate here that through this proceeding we obtain first
 closed expressions for the collective effects of charges, for which
 the perturbative expansion in small charge fugacities can be
 justified {\em a posteriori}. The various terms in this expansion can
 again be identified with vector charge classes. The leading terms
 correspond to the most relevant classes identified above and the
 stability region (shaded area) of Fig. \ref{fig.relev} remains
 unchanged.

\subsection{Screening}

 The renormalization of parameters on large scales due to fluctuations
 on small scales will now be calculated by integrating out pairs of
 particles from the partition sum following the approach of
 Kosterlitz.\cite{jmK74} Since our model possesses the vortex-charge
 duality, it is sufficient to perform the actual calculations only in
 one particle sector. The renormalization effects of the other sector
 are then obtained by a duality transformation. We adopt the results
 for the vortex sector directly from Ref. \CITE{S97}. However, we
 extend these previous calculations at two points. First, we calculate
 the flow equations for the fugacities to higher order in the
 fugacities. This extension is necessary to consider the parameter
 renormalizations not only within the phase with QLRO, but also in the
 phase where magnetic fields are relevant. This requires flow
 equations of higher order in the charge fugacities, which we can
 obtain through the duality mapping only from the terms of higher
 order in the vortex fugacities. Second, when the vortex dipoles are
 integrated out in the presence of charges, they renormalize the
 interactions among the charges (and vice versa). This feature was
 naturally absent in Ref. \CITE{S97}, where fields were not included.

 For technical convenience we allow replica vortices to take positions
 in the continuous space and not only on the lattice. The lattice
 provides only a cutoff (taken to be unity) for the minimal distance
 between vortices. Our present goal is to integrate out all
 configurations in the partition sum, where the smallest distance
 between two replica vortices (of any type) lies in the range $1 \leq
 |{\bf R}_1- {\bf R}_2| < 1+dl$. We follow Young's treatment of a
 vector Coulomb gas,\cite{apY79} which can be applied directly to our
 vectors ${\bf N}$. During this procedure two cases have to be
 distinguished: (i) the two close vortices are of opposite type, i.e.,
 ${\bf N}_1 + {\bf N}_2=0$ (in the continuous space subscripts to
 ${\bf N}$ denote the particle label instead of the position) and (ii)
 these two vortices are not of opposite type, i.e., ${\bf N}_1 +{\bf
 N}_2 \neq 0$.

 In case (i), where the pair of close vortices is composed
 of a vortex and its antivortex, they essentially annihilate. The
 leading interaction with the other particles, which are approximately
 considered to be far away, is through the dipole moment of the
 pair. Because of the polarizability of this dipole moment the pair
 screens the interaction between the other particles.\cite{KT73,jmK74}
 The screening of the interaction among vortices can be captured by a
 flow of the coupling
\begin{equation} 
\left. \frac {d}{dl} \right|_{\rm scr} {\bf K}_{\rm v} = 
2 \pi^3 {\bf K}_{\rm v} \cdot 
{\bf C}_{\rm v} \cdot {\bf K}_{\rm v},
\end{equation}
 where
\begin{equation}
\label{def.C}
C_{\rm v}^{ab} \equiv - \sum_{{\bf N} \neq {\bf 0}} 
N^a N^b y_{\rm v}^2({\bf N})
\end{equation}
 is the replica vortex density correlation at the cutoff distance [see
 Eq. (13) of Ref. \CITE{S97}]. Due to the angular coupling between
 vortices and charges there is an additional screening of the
 interaction among charges (see Appendix \ref{app.screen}):
\begin{equation} 
\label{Kc.screen}
\left. \frac {d}{dl} \right|_{\rm scr} {\bf K}_{\rm c} = 
- \frac {\pi}2 p^2 {\bf C}_{\rm v} .
\end{equation}
 Besides the screening of these coupling matrices, the annihilation of
 such dipoles changes the free energy per unit area and per replica by
\begin{equation}
\label{flow.f}
\left. \frac{d}{dl}  \right|_{\rm scr} {\cal F} = 
- \frac{\pi}n  \sum_{{\bf N} \neq {\bf 0}}  y_{\rm v}^2({\bf N}).
\end{equation}

 In case (ii) the two close vortices appear (in the eyes of the other
 particles, which are assumed to be far away) essentially as a single
 vortex of type ${\bf N}_3={\bf N}_1 +{\bf N}_2$, i.e. the dominant
 interaction with the other particles is not given by the dipole
 moment of the pair but by the total vorticity. Therefore we can
 ``replace'' the two vortices by the single one, i.e., we transfer the
 statistical weight of the entire particle configuration to a
 different particle configuration, where the two close vortices 1 and
 2 are replaced by a vortex of type 3 and all other particles are
 unchanged. By analogy with Young's analysis\cite{apY79} [leading to
 his Eq. (75)] we obtain
\begin{equation}
\label{monopole}
\left. \frac{d}{dl} \right|_{\rm scr} y_{\rm v}({\bf N}) = 
\pi \sum_{{\bf N}_1,{\bf N}_2 \neq {\bf 0}} 
\delta_{{\bf N},{\bf N}_1+{\bf N}_2} 
y_{\rm v}({\bf N}_1) y_{\rm v}({\bf N}_2).
\end{equation}
 The neutrality conditions (\ref{neutr}) are preserved under such
 types of recombination.

\subsection{Flow equations}

 In order to establish the RG flow equations, we combine the
 contributions from scaling and screening. Using the duality relations
 (\ref{dual.rep}) we infer the contributions by charge fluctuations
 immediately from those by vortex fluctuations that have been given
 above. Thus, after collecting all terms, we find the complete flow
 equations for the replicated system
\begin{mathletters}
\label{flow.y}
\begin{eqnarray}
\frac{d}{dl} y_{\rm v}({\bf N}) &=& 
\left(2- \pi {\bf N} \cdot {\bf K}_{\rm v} \cdot {\bf N}\right)
y_{\rm v}({\bf N}) 
\nonumber \\
&& + \pi \sum_{{\bf N}_1,{\bf N}_2 \neq {\bf 0}}
\delta_{{\bf N},{\bf N}_1+{\bf N}_2} 
y_{\rm v}({\bf N}_1) y_{\rm v}({\bf N}_2),
\\
\frac{d}{dl} y_{\rm c}({\bf n}) &=& 
\left(2- \pi {\bf n} \cdot {\bf K}_{\rm c} \cdot {\bf n}\right)
y_{\rm c}({\bf n})  
\nonumber \\
&&+ \pi \sum_{{\bf n}_1,{\bf n}_2 \neq {\bf 0}}
\delta_{{\bf n},{\bf n}_1+{\bf n}_2} 
y_{\rm c}({\bf n}_1) y_{\rm c}({\bf n}_2) ,
\end{eqnarray}
\end{mathletters}
 and 
\begin{mathletters}
\label{flow.param}
\begin{eqnarray}
\frac {d}{dl} {\bf K}_{\rm v} &=&  
2 \pi^3 {\bf K}_{\rm v} \cdot {\bf C}_{\rm v} \cdot {\bf K}_{\rm v} 
- \frac {\pi}2 p^2 {\bf C}_{\rm c} ,
\\
\frac {d}{dl} {\bf K}_{\rm c} &=&  
2 \pi^3 {\bf K}_{\rm c} \cdot {\bf C}_{\rm c} \cdot {\bf K}_{\rm c} 
- \frac {\pi}2 p^2 {\bf C}_{\rm v} ,
\\
\frac{d}{dl} {\cal F} &=& 
- \frac{\pi} n  \sum_{{\bf N} \neq {\bf 0}}  y_{\rm v}^2({\bf N}) 
- \frac{\pi} n  \sum_{{\bf n} \neq {\bf 0}}  y_{\rm c}^2({\bf n}).
\end{eqnarray}
\end{mathletters}
 Since we establish the flow equations for an infinitesimal change of
 the cutoff, we assume that we can simply add vortex and charge
 contributions, i.e., that there are no terms mixing vortex and charge
 fugacities. Such a mixing could arise in principle if one implies a
 minimal cutoff distance not only among vortices and among charges,
 but also between vortices and charges. We are inclined to believe
 that a hard-core interaction between vortices and charges is not
 necessary, because we do not expect divergences related to short
 distances of that type: due to the imaginary-valued nature of angular
 coupling (\ref{H.vc}) in the Hamiltonian it seems plausible that
 configurations with close vortex-charge pairs give no essential
 contribution to the partition sum.

 Since not only the original couplings ${\bf K}$ but also the particle
 density correlations ${\bf C}$ are replica-symmetric, this symmetry
 holds also for the screened couplings. Even more, the renormalized
 matrices ${\bf K}_{\rm v}$ and ${\bf K}_{\rm c}$ can be reexpressed
 in terms of renormalized parameters $K$ and $\sigma^\Pi$ preserving
 relations (\ref{E.K}a,b): Equations (\ref{flow.param}a,b) are
 equivalent to
\begin{mathletters}
\label{flow.par}
\begin{eqnarray}
\frac{d}{dl} K^{-1} &=& - 2 \pi^3 C_{\rm v, con} + 
\frac {\pi}2 p^2 K^{-2} C_{\rm c, con} ,
\\
\frac{d}{dl} \sigma^\Pi &=& -2 \pi^3 C_{\rm v, dis} + 
\frac {\pi}2 p^2 K^{-2}  [C_{\rm c, dis}
\nonumber \\
&& + \sigma^\Pi K (2 + n \sigma^\Pi K) 
(C_{\rm c, con}+ n C_{\rm c,dis})].
\end{eqnarray}
\end{mathletters}
 Thereby the particle correlation matrices were decomposed into the
 ``connected'' and ``disconnected'' contributions according to
\begin{equation}
C^{ab}=C_{\rm con} \delta^{ab} + C_{\rm dis}
\end{equation}
 for both vortices and charges.

\subsection{Replica limit $n \to 0$}

 So far, the derivation of the RG flow equations for the replicated
 system have been conceptually straightforward. The less trivial part
 is to take the replica limit, i.e., to perform an analytic
 continuation for $n \to 0$.

 The first step in this direction was the reduction of the ${\bf K}$
 matrix flow equation to the parameter flow equations
 (\ref{flow.par}). We focus on these equations, since the physical
 large-scale properties of our system show up in the flow of these
 parameters. In order to send $n \to 0$ in these flow equations, we
 have to overcome only one hurdle: to make $n$ an explicit parameter
 in the connected and disconnected correlations. Here again we follow
 the route of Ref. \CITE{S97}, where the problem was solved for the
 vortex correlations, and treat charges on the same footing.

 One easily realizes from the flow equations for the fugacities that
 large vortices (charges) with $|N^a|>1$ ($|n^a|>1$) are less relevant
 than small vortices (charges) with $|N^a| \leq 1$ ($|n^a| \leq
 1$). Therefore it is a good approximation to restrict the further
 calculation to the small particles (this approximation is made for
 simplicity, the inclusion of large particles is possible). The
 correlations can then explicitly be evaluated after introducing a
 auxiliary random variable ${\cal A}$. Its distribution is defined to
 be Gaussian with average $[{\cal A}]_{\cal A}=0$ and variance $[{\cal
 A}^2]_{\cal A}=1/2$ (averages over ${\cal A}$ are denoted by
 $[\dots]_{\cal A}$). We decompose the core energy matrices
 (\ref{E.K}c,d) into
\begin{equation}
E^{ab}=E \delta^{ab} - \hat E
\end{equation}
 for vortices and charges (one specifies by adding the corresponding
 subscripts ``v'' or ``c''). The initial values of $E$ given in
 Eq. (\ref{defs.E}) and
\begin{equation}
\hat E_{\rm v}= \frac{\pi \gamma \sigma^T K^2}{1+n \sigma^T K}, \ \ \ 
\hat E_{\rm c}= - \frac{\gamma p^2 \sigma^L}{4\pi}.
\end{equation}
 Next, we introduce weights (proportional to the fugacities $e^{-2E}$
 of the elementary particles)
\begin{equation}
\label{def.z}
z_\pm \equiv e^{-2(E \pm {\cal A} \sqrt{2 \hat E})}, \ \ \
z \equiv 1+ z_+ + z_- ,
\end{equation}
 and, combining Eqs. (\ref{fuga}) with (\ref{def.C}) and its dual
 counterpart, we rewrite the correlations as
\begin{mathletters}
\begin{eqnarray}
C^{aa}&=& - \left[(z_+ + z_-) z^{n-1} \right]_{\cal A},
\\
C^{ab}&=& - \left[[(z_+ - z_-)^2 z^{n-2} \right]_{\cal A},
\end{eqnarray}
\end{mathletters}
 where we take $a \neq b$. In this representation the number of
 replicas became an explicit variable that can be sent to zero,
 resulting in
\begin{mathletters}
\label{C.z}
\begin{eqnarray}
C_{\rm dis}&=& - \left[\left(\frac{z_+ - z_-}{1+ z_+ + z_-}\right)^2
\right]_{\cal A},
\\
C_{\rm con}&=& - \left[\frac{z_+ + z_- + 4z_+ z_-}{(1+ z_+ + z_- )^2}
\right]_{\cal A}.
\end{eqnarray}
\end{mathletters}
 Then the replica limit of the flow equations (\ref{flow.par}) for $K$
 and $\sigma^\Pi$ can be taken right away, since now all dependences
 on $n$ are explicit.

 The contributions to the flow of the free energy (\ref{flow.param}c)
 can be rewritten as\cite{S97}
\begin{equation}
\label{F.z}
\frac 1n \sum_{{\bf N} \neq {\bf 0}} y_{\rm v}^2({\bf N}) = \frac 1n
[z_{\rm v}^n-1]_{\cal A} \to [\ln z_{\rm v}]_{\cal A},
\end{equation}
 where in the last expression the replica limit has already been
 performed. The analogous charge contribution is obtained by duality
 substitutions.

\subsection{Fugacity expansion}

 Combining expressions (\ref{flow.param}c), (\ref{flow.par}),
 (\ref{C.z}), and (\ref{F.z}) the flow of the physical quantities $K$,
 $\sigma^\Pi$, and ${\cal F}$ can be expressed through the fugacities
 $z_\pm$ for both types of particles.  According to Eq. (\ref{defs.E})
 these fugacities are small for large $K$ and small $H$ and one might
 be tempted to simplify the expressions (\ref{C.z}) and (\ref{F.z}) by
 a truncated expansion in these fugacities.

 The early studies of the disordered $XY$
 model\cite{CO82,RSN83,GK,PK91} made use of this fugacity
 expansion. The invalidity of this expansion was recognized later on
 and was considered as a possible reason for the destruction of QLRO
 even by infinitesimally weak disorder.\cite{seK93,MW97} However, it
 was shown recently\cite{NSKL95,KN96,S97,lhT96} for the model without
 fields that the mere breakdown of the fugacity expansion does not
 imply the relevance of vortices and the destruction of QLRO. In other
 terms: in the vortex sector the closed expressions (\ref{C.z}) and
 (\ref{F.z}) are {\em finite} and they are to be evaluated
 nonperturbatively.  Therefore it is a central issue of the present
 work to reexamine the validity of the fugacity expansion for charges
 that underlies Refs. \CITE{CO82,PK91}.

 To be explicit let us examine the ``smallness'' of the charge
 fugacities $z_{{\rm c} \pm}$ defined in Eq. (\ref{def.z}). In
 contrast to the real-valued vortex fugacities they are complex-valued
 since $\sqrt{ \hat E_{\rm c}}$ is purely imaginary. Their absolute
 value
\begin{equation}
|z_{{\rm c} \pm}|= e^{-2E_{\rm c}} \approx e^{-1/H} \approx \frac{H_0^2}4 
\end{equation}
 is small for weak fields {\em independently} of the disorder
 represented by the variable ${\cal A}$. Therefore it is safe to
 expand the expressions correlations (\ref{C.z}) for charges {\em
 before} performing the average over ${\cal A}$. Again, this is in
 sharp contrast to the case of vortices, where the fugacities become
 arbitrarily large in the tails of the distribution of ${\cal
 A}$. Thus, the mechanism that led to the failure of the fugacity
 expansion for vortices is {\em not} effective for charges.

 Thus, we are entitled to use the fugacity expansion for charges to
 study the stability QLRO to weak fields.  We expand the charge
 correlations (\ref{C.z}) [and analogously the source of the flow of
 the free energy, the dual counterpart of Eq. (\ref{F.z})] to second
 order in the fugacities $z_{{\rm c} \pm}$ and perform the average
 over ${\cal A}$:
\begin{mathletters}
\label{cocol}
\begin{eqnarray}
C_{\rm c, dis} &=&  
2 y_{{\rm c}\uparrow \downarrow}^2 - 2 y_{{\rm c}\uparrow \uparrow}^2
+ \cdots,
\\
C_{\rm c, con} &=& 
-2 y_{{\rm c}\uparrow}^2 + 4 y_{{\rm c}\uparrow \uparrow}^2+ \cdots,
\\
{[ \ln z_{\rm c} ]}_{\cal A} &=& 
2 y_{{\rm c}\uparrow}^2 -y_{{\rm c}\uparrow \downarrow}^2 -
y_{{\rm c}\uparrow \uparrow}^2 + \cdots
\end{eqnarray}
\end{mathletters}
 We have replaced the terms in the series by vector charge fugacities
 using the identification (\ref{fuga}) and substituted the symbols
 $y_{\rm c}({\bf n})$ by the more illustrative arrow symbols. The
 initial values of these fugacities are given by
\begin{mathletters}
\label{y.c.0}
\begin{eqnarray}
y_{{\rm c}\uparrow} &=& e^{-[1/2H + 
(\gamma p^2/4 \pi) (K^{-1} + \sigma^L)]} ,
\\
y_{{\rm c}\uparrow \downarrow} &=& 
e^{-[1/H + (\gamma p^2/2 \pi)  K^{-1}]},
\\
y_{{\rm c}\uparrow \uparrow} &=& 
e^{-[1/H + (\gamma p^2/2 \pi) (K^{-1} + 2  \sigma^L)]} .
\end{eqnarray}
\end{mathletters}
 All bare charge fugacities are determined by only three physical
 parameters $K$, $\sigma^L$, and $H$. This relation is abandoned under
 renormalization, where the flow of $K$ and $\sigma^L$ follows
 Eq. (\ref{flow.par}) and where we have a whole set of flow equations
 (\ref{flow.y}) for the charge fugacities instead of a single flow
 equation for $H$. It is interesting to note that a fugacity with $s$
 arrows is of order $H_0^{s}$ in the original field [remember relation
 (\ref{def.f})]. This order $s=s_\uparrow + s_\downarrow$ is just the
 total number of positive/negative elementary charges introduced in
 Eq. (\ref{mo.re.c}) above.

 Since the fugacity expansion is valid for the charges, it is
 legitimate to analyze the stability of QLRO by demanding that all
 classes of vector charges have to be irrelevant. This analysis has
 been performed in Sec. \ref{sec.scal} leading to the shaded stability
 area in Fig. \ref{fig.relev}. As long as all charges are irrelevant,
 the second order terms in the flow equation (\ref{flow.y}b) play
 actually no role. For a {\em qualitative} study of the flow of charge
 fugacities slightly inside the region where charges are relevant, we
 keep only the two charge classes $y_{{\rm c}\uparrow}$ and $y_{{\rm
 c}\uparrow \downarrow}$. This should be a reasonable approximation
 since it is always one of these two classes that brings about the
 instability of QLRO at the low-temperature border of the stability
 region and that represent the most relevant charge class among all
 classes and for all parameters, as discussed after
 Eq. (\ref{mo.re.c}). Therefore we truncate the charge fugacity flow
 equations (\ref{flow.y}b) to
\begin{mathletters}
\begin{eqnarray}
\frac{d}{dl} y_{{\rm c}\uparrow} &=& 
[2 - (p^2 / 4 \pi) (K^{-1} + \sigma^L)]  y_{{\rm c}\uparrow} 
- 2 \pi y_{{\rm c}\uparrow}y_{{\rm c}\uparrow \downarrow} ,
\\
\frac{d}{dl} y_{{\rm c}\uparrow \downarrow} &=& 
[2 - (p^2 / 2 \pi) K^{-1}]  y_{{\rm c}\uparrow \downarrow} 
 + 2 \pi [y_{{\rm c}\uparrow}^2-2 y_{{\rm c}\uparrow \downarrow}^2] .
\end{eqnarray}
\end{mathletters}
 However, we stress that these truncated flow equations are
 incomplete, i.e., these two selected fugacities are coupled to all
 other fugacities in the complete flow equations. We have dropped also
 charge class $\uparrow \uparrow$, which also brings about only
 quantitative modifications. Only for $\sigma^L=0$ they are as
 relevant as charges $\uparrow \downarrow$ and should be kept, e.g., to
 obtain the vanishing of the disconnected correlation (\ref{cocol}a).

 Concerning the processing of the vortex sector, we actually don't go
 beyond Ref. \CITE{S97}. That is, we restrict our quantitative
 analysis to the region, where vortices are irrelevant and the
 second-order terms to the vortex fugacity flow equations
 (\ref{flow.y}a) can be discarded. The linear flow equations for all
 $y_{\rm v}({\bf N})$ can be reduced to flow equations for $E_{\rm v}$
 and $\hat E_{\rm v}$.\cite{S97} Avoiding a fugacity expansion, the
 vortex correlations $C_{\rm v}$ can be evaluated asymptotically for
 large length scales and can be expressed in terms of an {\em
 effective} vortex fugacity $y_{\rm v}$,
\begin{mathletters}
\begin{eqnarray}
C_{\rm v, con} &=& -2 \tau^* y_{\rm v}^2 ,
\\
C_{\rm v, dis} &=& -2 (1-\tau^*) y_{\rm v}^2 ,
\\
{[ \ln z_{\rm v} ]}_{\cal A} &=& \frac 2 {\tau^*},
\end{eqnarray}
\end{mathletters}
 with a fraction of ``polarizable dipoles''
\begin{equation}
\label{def.tau}
\tau^* \equiv \min(1,1/2\sigma^T K).
\end{equation}
 This fraction is unity for temperatures $K^{-1} > 2 \sigma^T$ and
 goes to zero as the temperature drops below $K^{-1} < 2 \sigma^T$,
 where vortices start to freeze. The initial value and the
 renormalization flow of the effective fugacity is given by\cite{S97}
\begin{mathletters}
\label{y.v}
\begin{eqnarray}
y_{\rm v}^2&=&
\frac 12\left[\frac{z_+ + z_-}{1+ z_+ + z_-} \right]_{\cal A},
\\
\frac{d}{dl} y_{\rm v} &=& 
[2-\pi \tau^* K(1-\sigma^T \tau^* K)]y_{\rm v}.
\end{eqnarray}
\end{mathletters}

 To summarize the above derivations we collect the full set of flow
 equations that constitute one of our main findings:
\begin{mathletters}
\label{flow.full}
\begin{eqnarray}
\frac{d}{dl} K^{-1} &=& 4 \pi^3 \tau^* y_{\rm v}^2 
- \pi p^2 K^{-2} y_{{\rm c}\uparrow}^2,
\\
\frac{d}{dl} \sigma^\Pi &=& 4 \pi^3 (1-\tau^*) y_{\rm v}^2 
-\pi p^2 [ 2 \sigma^\Pi K^{-1} y_{{\rm c}\uparrow}^2
\nonumber \\
&& -K^{-2} y_{{\rm c}\uparrow \downarrow}^2],
\\
\frac{d}{dl} y_{\rm v} &=&  [2-\pi \tau^* K(1-\sigma^T \tau^* K)]
y_{\rm v} ,
\\ 
\frac{d}{dl} y_{{\rm c}\uparrow} &=& 
[2 -(p^2 / 4 \pi) (K^{-1} + \sigma^L)]  y_{{\rm c}\uparrow} 
- 2 \pi y_{{\rm c}\uparrow}y_{{\rm c}\uparrow \downarrow} ,
\\
\frac{d}{dl} y_{{\rm c}\uparrow \downarrow} &=& 
[2 - (p^2 / 2 \pi) K^{-1}]  y_{{\rm c}\uparrow \downarrow} 
 + 2 \pi [y_{{\rm c}\uparrow}^2-2 y_{{\rm c}\uparrow \downarrow}^2] ,
\\
\frac{d}{dl} {\cal F} &=&
-\frac{2 \pi}{\tau^*} y_{\rm v}^2 - \pi ( 2 y_{{\rm c}\uparrow}^2  
- y_{{\rm c}\uparrow \downarrow}^2).
\end{eqnarray}
\end{mathletters}
 This set of flow equations has no longer an explicit duality between
 vortices and charges (even if we ignore the second-order terms in the
 flow equations of the charge fugacities), because the approximations
 used to treat the coupling of vortices and charges to disorder have
 to be taken in a fundamentally different way, mainly because of the
 validity (invalidity) of the fugacity expansion for charges
 (vortices). In particular it was possible to retain only two charge
 classes, whereas the collective effect of all vortex classes is
 represented by the effective vortex fugacity and the appearance of
 the parameter $\tau^*$ (that actually parametrizes the most relevant
 vortex class).

 However, if we specialize flow equations (\ref{flow.full}) to the
 case without disorder, the duality reappears and we retrieve the flow
 equations of Ref. \CITE{JKKN77}.\cite{note.comp.JKKN77} If we switch
 off the field, i.e., if we drop the charge sector, the flow equations
 (\ref{flow.full}) are reduced per construction exactly to those
 derived recently.\cite{S97,lhT96} In the presence of random fields
 [where $\sigma^L =\infty$ and $y_{{\rm c} \uparrow}=0$, see
 Eq. (\ref{y.c.0})], the flow equations reduce in the charge sector to
 those of CO.\cite{note.comp.CO82} However, the present flow equations
 allow also the consideration of finite $\sigma^L$. The charge sector
 for uniform fields has structural similarities with the flow
 equations that have been given (but that have not been evaluated) in
 Ref. \CITE{HG97}.\cite{note.comp.HG97} The flow equations in
 Ref. \CITE{PK91} miss the contributions from fugacities $y_{{\rm c}
 \uparrow \downarrow}$ and the freezing of vortices for $\tau^*<1$.

\section{Results}

 We now determine the large-scale properties of the model by
 integrating the flow equations (\ref{flow.full}). Thereby the initial
 parameters $K$ and $\sigma^\Pi$ are mapped for $l= \infty $ onto
 $K_\infty$ and $\sigma^\Pi_\infty$. The fugacities, which are
 initially given by Eqs. (\ref{y.c.0}) and (\ref{y.v}a), flow
 independently. Note that $\tau^*$ is also a flowing parameter which
 is determined by the flowing $K$ and $\sigma^T$ through relation
 (\ref{def.tau}).

 From the flow equations one easily reads off how vortex and charge
 fluctuations renormalize the physical parameters qualitatively.
 Equation (\ref{flow.full}a) means that the effective temperature is
 increased by vortices but reduced by charges, corresponding to
 disordering tendency of vortices and the ordering tendency of
 charges.  Vortices always increase the strength of disorder but the
 effect of charges is ambivalent in Eq. (\ref{flow.full}b). The
 competition between the various terms leads to a rich structure of
 the phase diagram. For the clarity of the analysis we first break the
 problem into parts that we eventually assemble to the complete
 picture.

\subsection{Vortex matter}

\begin{figure}
\epsfxsize=0.9 \linewidth
\epsfbox{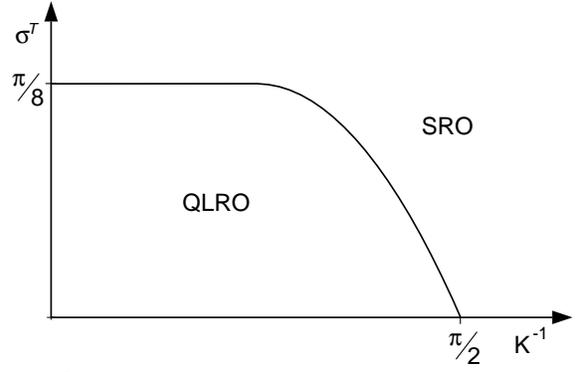}
\narrowtext
\caption{
 Nonreentrant phase diagram in the absence of fields according to
 Refs. \protect\CITE{NSKL95,CF95,KN96,S97,lhT96} and as it is
 reproduced by the flow equations (\protect\ref{flow.full}).}
\label{fig.pdg.v}
\end{figure}

 For completeness and further reference we recall the main properties
 of the system in the absence of fields:\cite{NSKL95,S97,lhT96} At
 temperatures below the KT transition temperature of the pure system,
 that is, $K^{-1}_\infty \leq \pi/2$, vortices are irrelevant even in
 the presence of random phase shifts with a strength below the
 critical value (see Fig. \ref{fig.pdg.v})
\begin{equation}
\label{irrelev.v}
\sigma^T_\infty \leq \left\{
\begin{array}{ll}
\frac 1 {K_\infty} - \frac 2{\pi K_\infty^2} & \mbox{for} \ \ 
\frac {\pi}4 \leq K^{-1}_\infty \leq \frac{\pi}2,
\\
\frac{\pi}8 & \mbox{for} \ \ K^{-1}_\infty \leq \frac{\pi}4.
\end{array}
\right.
\end{equation}
 The increase of $K^{-1}$ and $\sigma^T$ under renormalization implies
 that the phase with QLRO is less extended in the $(K^{-1},\sigma^T)$
 plane of the unrenormalized parameters as compared to the
 $(K^{-1}_\infty ,\sigma^T_\infty)$ plane of the renormalized
 parameters. Nevertheless, this shrinking is only a quantitative
 modification and QLRO is stable for low temperatures and weak
 disorder (see Fig. 3 of Ref. \CITE{S97}).

 It is evident from the dependence of the flow equations on the
 parameter $\tau^*$ that the physics within the phase with QLRO is
 different for high temperatures and for low temperatures. Vortices
 start to freeze and the breakdown of the fugacity expansion sets in
 at temperatures below the line $K_\infty^{-1} = 2 \sigma^T_\infty$
 separating these two regimes.

 The spin correlation function $\Gamma_p$ decays algebraically as in
 the spin wave approximation (\ref{Gamma.sw}) but with a renormalized
 exponent
\begin{equation}
\label{eta.inf}
\eta_\infty =\frac {p^2}{2\pi} 
\left(\frac{1}{K_\infty} +\sigma^L_\infty\right).
\end{equation}
 Through the renormalization of $K$ and $\sigma^L$ the value of
 $\eta_\infty$ depends implicitly also on the transverse disorder.
 Its value is $\eta_\infty=p^2/4$ at the high-temperature end of the
 transition line and drops continuously down to $\eta_\infty=p^2/16$
 at the zero-temperature end of the transition line. The exponent
\begin{equation}
\label{overeta.inf}
\overline{\eta}_\infty = \frac 1{\pi K_\infty}
\end{equation}
 of the Edwards-Anderson spin correlation (\ref{EA.corr}) assumes
 values in the range $0 \leq \overline{\eta} \leq 2$ and also varies
 nonuniversally along the QLRO/SRO transition. When the transition
 line is approached from the high-temperature side, the correlation
 length diverges exponentially (with exception of a special point at
 $K_\infty^{-1} = 2 \sigma^T_\infty$\cite{lhT96}) with decreasing
 distance from the transition line as in the absence of
 disorder.\cite{S97,lhT96} Within the QLRO phase glassy features are
 found in vortex density correlations for temperatures $K_\infty^{-1}
 \leq 2 \sigma^T_\infty$.\cite{S97}

\subsection{Charge matter}

 Let us now consider the complementary case where we evaluate the
 effect of fields in the absence of vortices. Taking into account
 renormalization effects, the linearized charge-fugacity flow
 equations indicate the irrelevance of fields at high enough
 temperature even for arbitrarily strong disorder:
\begin{equation}
\label{irrelev.c}
K^{-1}_\infty \geq \left\{
\begin{array}{ll}
\frac{8 \pi}{p^2} -\sigma^L_\infty  & \mbox{for} \ \ 
\sigma^L_\infty \leq  \frac {4 \pi}{p^2},
\\
\frac{4 \pi}{p^2} & \mbox{for} \ \ \sigma^L_\infty \geq \frac{4 \pi}{p^2}.
\end{array}
\right.
\end{equation}
 In the low-temperature region of the phase diagram, where the
 magnetic field is relevant, we can determine the critical exponent
 $\delta$ that defines the scaling relation $M \sim H_0^{1/\delta}$
 between the magnetization $M$ and the field $H_0$. From the
 linearized flow equations of the charge fugacities we determine the
 exponent $x$ in the scaling relation $H_0 \sim L^x$ between the field
 amplitude and the system size $L \equiv e^l$. From the two charge
 classes under consideration we find
\begin{mathletters}
\begin{eqnarray}
y_{{\rm c} \uparrow} &\sim &H_0 
\sim L^{2-(p^2/4 \pi)(K^{-1} + \sigma^L)},
\\
y_{{\rm c} \uparrow \downarrow} &\sim& H_0^2 
\sim L^{2-(p^2/2 \pi) K^{-1}}.
\end{eqnarray}
\end{mathletters}
 From these two scaling relations one obtains at first sight two
 different values for $x$. The actual value of $x$, which describes
 the relevance of $H_0$, is given by the larger of the two
 values. Therefore we find
\begin{equation}
\label{x}
x=\left\{
\begin{array}{ll}
2-\frac{p^2}{4 \pi}(K^{-1} + \sigma^L) &
\mbox{for} \ \ \sigma^L \leq \frac{4 \pi}{p^2},
\\
1-\frac{p^2}{4 \pi}K^{-1} &
\mbox{for} \ \ \sigma^L \geq \frac{4 \pi}{p^2},
\end{array}
\right.
\end{equation}
 as special cases of Eq. (\ref{mo.re.c}). To determine this exponent
 it is important to keep in mind that different vector charges are of
 different order in the field. At this point one might question again
 whether charge classes different from the two considered ones could
 change the value of $x$. This is not the case and can be understood
 from an inspection of the $x$ values for the other classes. These
 values are obtained by dividing the exponent in Eq. (\ref{mo.re.c})
 by the power $s=s_\uparrow+s_\downarrow$ of the scaling relation
 between the fugacity and the field. The different order of the charge
 fugacities in the field is also the reason that the separation line
 $K^{-1} =\sigma^L$, where dominance switches between charges of class
 $\uparrow$ or $\uparrow \downarrow$, does not coincide with the
 separation line $\sigma^L=4 \pi/p^2$ where the exponent $x$ has its
 nonanalytic dependence on $K^{-1}$ and $\sigma^L$.

 From the assumption that the magnetization yields an extensive
 contribution to the free energy, $\Delta F \sim L^2 \sim H_0^{2/x}$
 and the definition of the magnetization $M \sim \partial \Delta
 F/\partial H_0$ we identify the exponent
\begin{equation}
\label{delta.inf}
\delta=\frac x{2-x}= \left\{
\begin{array}{ll}
\frac 4 {\eta_\infty} -1 &
\mbox{for} \ \ \sigma^L_\infty \leq \frac{4 \pi}{p^2},
\\
\frac{1- p^2/ 4 \pi K_\infty}{1+ p^2/ 4 \pi K_\infty} &
\mbox{for} \ \ \sigma^L_\infty \geq \frac{4 \pi}{p^2}.
\end{array}
\right.
\end{equation}
 This expression reproduces the universal value $\delta=15$ for fields
 with $p=1$ at the KT transition of the pure system.\cite{jmK74}

\begin{figure}
\epsfxsize=0.9 \linewidth
\epsfbox{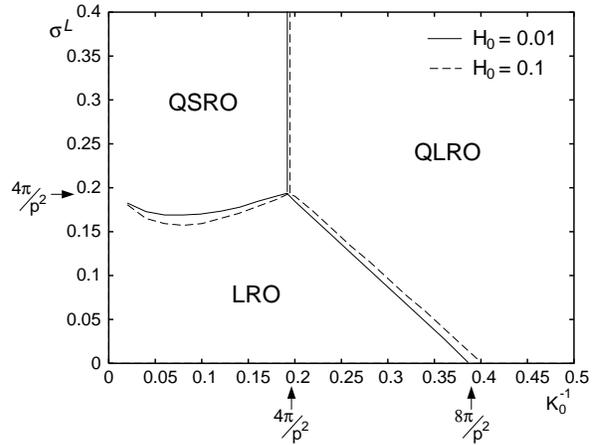}
\narrowtext
\caption{
 Phase diagram in the absence of vortices. The transition lines
 between the phases LRO, QLRO, and QSRO depend only weakly on small
 fields $H_0\ll 1$. The transition between LRO and QSRO is almost
 horizontal for weak fields.}
\label{fig.pdg.c}
\end{figure}

 At low temperatures, where the field is relevant, we find two
 different phases. For weak disorder $\sigma^L \lesssim 4\pi / p^2$
 the ordering tendency of the fields dominates over the disorder and
 LRO is established. In the opposite case $\sigma^L \gtrsim 4\pi /
 p^2$ the disorder wins over the field and results in
 quasi-short-range order (QSRO) characterized below. The phase
 diagram, obtained from a numerical integration of the flow equations
 in the absence of vortices, is depicted in Fig. \ref{fig.pdg.c}.

 LRO is found in a region $\sigma^L \lesssim 4 \pi /p^2$ and $K_0^{-1}
 \lesssim 8 \pi /p^2 - \sigma^L$ of the unrenormalized parameters.
 Under renormalization both charge fugacities saturate at a finite
 value and drive these parameters to zero,
\begin{mathletters}
\begin{eqnarray}
y_{{\rm c} \uparrow} &\to& \frac 1 \pi,
\\
y_{{\rm c} \uparrow \downarrow} &\to& \frac 1 \pi,
\\
K^{-1} &\approx& \frac{\pi}{p^2 l},
\\
\sigma^L &\approx& \frac{\pi}{p^2 l}.
\end{eqnarray}
\end{mathletters}
 On large scales $K^{-1}$ and $\sigma^L$ decay only inversely
 proportional to the logarithm of the length scale, and so does the
 effective exponent $\eta \approx 1/l$. Therefore this fixed point
 corresponds indeed to a phase with LRO, where
\begin{equation}
\Gamma_p ({\bf r}) \to {\rm const.} >0 
\end{equation}
 for $r \to \infty$. The fact that the fugacities saturate at a finite
 value $1/ \pi$ is consistent with a very (maybe infinitely) strong
 field $H_0$ at the fixed point, since even an infinitely strong
 initial field would lead to finite fugacities, see Eq. (\ref{y.c.0}).
 The analysis of the LRO phase cannot be taken quantitatively, since
 the flow of $K^{-1}$ and $\sigma^L$ entails the relevance of other
 charge classes that have been ignored in this analysis. However, we
 expect modifications to be only quantitative since the omitted
 charges are less relevant than the retained charges [see the
 discussion near Eq. (\ref{mo.re.c})].

 QSRO is found in a region $\sigma^L \gtrsim 4 \pi /p^2$ and $K_0^{-1}
 \lesssim 4 \pi /p^2$ of the unrenormalized parameters and identified
 with the following asymptotic flow:
\begin{mathletters}
\label{fix.QSRO}
\begin{eqnarray}
y_{{\rm c} \uparrow} &\to& 0,
\\
y_{{\rm c} \uparrow \downarrow} &\to& \frac 1{2 \pi} \left(
1-\frac{p^2}{4 \pi K_\infty} \right),
\\
K^{-1} &\to& K^{-1}_\infty \leq \frac {4 \pi}{p^2},
\\
\sigma^\Pi &\approx& \frac{p^2}{4 \pi K_\infty^2} \left(
1-\frac{p^2}{4 \pi K_\infty} \right)^2 l.
\end{eqnarray}
\end{mathletters}
 Since both $\sigma^\Pi$ increase without bounds, there is no fixed
 point in a strict sense. According to our analysis in Sec.
 \ref{sec.scal} {\em all} charge classes different from $\uparrow
 \downarrow$ become actually irrelevant at temperature not too far
 below $K^{-1}_\infty=4 \pi /p^2$. However, at significantly lower
 temperatures other classes will be relevant, e.g., $\uparrow \uparrow
 \downarrow \downarrow$ at $K^{-1}_\infty=2 \pi /p^2$. The divergence
 of $\sigma^L$ means that the system behaves asymptotically like in
 the presence of {\em random} fields. Since both $\sigma^\Pi$ don't
 couple back into the flow equations, their effect on the angle
 difference correlation function can be evaluated in the spin-wave
 approximation accounting for the scale dependence of $\sigma^L$,
\begin{equation}
\overline{\langle (\theta_{\bf r} - \theta_{\bf 0})^2 \rangle}
\approx \frac 12 \left( \frac{p}{2 \pi K_\infty^2} \right)^2 
\left( 1-\frac{p^2}{4 \pi K_\infty^2} \right)^2
\ln^2 r ,
\end{equation}
 as found in Refs. \CITE{CO82,GH82,VF84}. In the context of crystals
 grown on disordered substrates this behavior of the correlation
 function is called ``super-rough''.\cite{TV90} To the extent that
 only the class $\uparrow \downarrow$ contributes to the
 renormalization, $K$ is not renormalized. For the random field model
 this non-renormalization can be shown in general on the basis of a
 statistical symmetry.\cite{CO82,GS85} Then the amplitude of the
 difference correlation in front of $\ln^2 r$ is universal in terms of
 the transition temperature $T_{\uparrow \downarrow}$, which is
 determined by $K_\infty ^{-1}=4\pi / p^2$, namely $\overline{\langle
 (\theta_{\bf r} - \theta_{\bf 0})^2 \rangle} \approx
 (2/p^2)(1-T/T_{\uparrow \downarrow})^2 \ln^2 r $ for $T \lesssim
 T_{\uparrow \downarrow}$ (Our quantitative analysis has to be
 restricted to parameters, where the fixed-point value of $y_{\uparrow
 \downarrow}$ is small. Otherwise higher orders in this fugacity can
 no longer be neglected). Our calculation agrees in the prefactor of
 $\ln^2r$ with that of Carpentier and Le Doussal.\cite{CD97} The
 increase of the difference correlation faster than logarithmically
 but slower than linearly in the length implies a faster than
 algebraic but slower than exponential decay of the spin correlation,
\begin{equation}
\Gamma_p(r) \sim r^{- (1-T/T_c)^2 \ln r}, 
\end{equation}
 i.e. the correlation length is still infinitely large. To emphasize
 this we call the order quasi-short-ranged.

 In the QSRO phase the competition between the spin coupling energy
 and the magnetization energy leads to a state, which has less order
 than each of the contributions alone would have in the absence of the
 other contribution (QLRO or LRO, respectively). This competition
 results in a frustration typical for glassy disordered systems. Since
 the symmetry-breaking field does not induce a finite magnetization,
 the scaling relation and the exponent $\delta$ are actually
 meaningless in this phase.

 The QSRO phase can be considered as ``glass'' since the disorder
 ($\sigma^L$) dominates the physics on large scales. In addition, the
 dominant fluctuations are vector charges of type $\uparrow
 \downarrow$, i.e. fluctuations that are correlated in different
 replicas and usually would imply a finite Edwards-Anderson order
 parameter. However, we are not aware of such an order parameter in
 the strict sense for this phase. But we observe that the
 Edwards-Anderson correlation (\ref{EA.corr}) decays only
 algebraically with finite ${\overline \eta}_\infty$, whereas the spin
 correlation $\Gamma_p$ decays faster than algebraically. This
 indicates that the conformations of $\theta$ are disorder dominated
 and that thermal fluctuations are not stronger than in the absence of
 disorder. Additional evidence for the glassy nature of this phase can
 be derived from dynamical properties,\cite{S95} which are beyond the
 scope of the present work.
 
 The transitions between the three phases LRO, QLRO and QSRO are of
 different nature. When the LRO/QLRO transition is approached from
 above, the spin correlation exponent reaches the universal value
 $\eta_\infty=4$,\cite{PK91} and $\delta=0$ when it is reached from
 below since the field becomes irrelevant. These values are thus
 independent of weak disorder with $\sigma^L_\infty \leq
 4\pi/p^2$. Along the QSRO/QLRO transition the exponent $\eta_\infty
 \geq 4$ has a nonuniversal dependence on the disorder strength, and
 $\delta$ is not a meaningful quantity. However,
 $\overline{\eta}_\infty=4/p^2$ is universal along this
 line.\cite{CO82}

 The phase transition LRO/QSRO is located near $\sigma^L_\infty =
 4\pi/p^2$, see Fig \ref{fig.pdg.c}. The transition line found from
 the numerical integration of the flow equations is bent towards
 smaller $\sigma^L$ and the bending decreases with decreasing field
 amplitude. In Eq. (\ref{x}) we have found the scaling exponent of the
 field to change along the line $\sigma^L_\infty = 4\pi/p^2$, which
 indicated that the instability of QLRO against the field differs
 above and below this line. However, a change of the exponent for
 initial scaling of the field does not necessarily imply that the
 renormalization flow drifts to different fixed points. We can provide
 an additional argument in support of this location of the transition
 line. Let us identify the difference between the phases with LRO or
 QSRO by the presence or absence of $y_{{\rm c} \uparrow}$ on large
 scales. We further consider very weak fields such that $K$ and
 $\sigma^L$ are very weakly renormalized (on intermediate
 scales). Then we may use an adiabatic approximation for the charge
 fugacities, i.e., we suppose the values of the fugacities to be
 determined through $d y_{{\rm c} \uparrow}/dl=0$ and $d y_{{\rm c}
 \uparrow \downarrow}/dl=0$ as a function of slowly varying $K^{-1}$
 and $\sigma^L$. From the flow equations (\ref{flow.full}) one obtains
 two different solutions
\begin{mathletters}
\begin{eqnarray}
\mbox{LRO} & \quad &
\left\{
\begin{array}{l}
y_{{\rm c} \uparrow \downarrow} \approx \frac 1 {2\pi}
\left[2-\frac{p^2}{4\pi}\left(\frac 1K + \sigma^L\right)\right],
\\
y_{{\rm c} \uparrow}^2 \approx \frac  {y_{{\rm c} \uparrow
\downarrow}}\pi
\left[1-\frac{p^2 \sigma^L}{4\pi}\right],
\end{array}
\right.
\\
\mbox{QSRO} & \quad &
\left\{
\begin{array}{l}
y_{{\rm c} \uparrow \downarrow} \approx \frac 1 {2\pi}
\left[1-\frac{p^2}{4\pi K}\right],
\\
y_{{\rm c} \uparrow}=0,
\end{array}
\right.
\end{eqnarray}
\end{mathletters}
 which are precursors of the LRO and QSRO fixed points. Eventually
 these precursors flow towards the true fixed points because of the
 scale dependence of $K$ and $\sigma^L$. These two solutions merge
 again at the line $\sigma^L = 4\pi/p^2$.  From a linear stability
 analysis of the fugacity flow equations one finds that the LRO
 precursor is stable and the QSRO precursor is unstable for $\sigma^L
 < 4\pi/p^2$ and the QSRO precursor is stable for $\sigma^L >
 4\pi/p^2$. The LRO precursor is unphysical for $\sigma^L > 4\pi/p^2$
 since there $y_{{\rm c} \uparrow}^2<0$.  Although this analysis
 supports the horizontal location of the LRO/QSRO phase boundary in
 the limit of weak fields, we should like to remind again that the
 transition can be described only qualitatively since this analysis
 includes only two charge types and the other neglected types are
 expected to give quantitative corrections even to the borders of the
 LRO phase, where the fixed point values of the charge fugacities
 remain finite.

 It is worthwhile to point out that the irrelevance of fields in the
 present context is equivalent to the irrelevance of the periodic
 crystalline potential for the roughening transition of crystals with
 correlated substrate disorder (the mapping between these two problems
 is the one discussed in Sec. \ref{sec.map}, $\phi$ corresponds to
 the substrate conformation). The latter problem has been examined in
 Ref. \CITE{S95} and the phase diagram obtained there (Fig. 2 therein)
 is the analog to the onset of relevance of fields in our
 Fig. \ref{fig.relev}. In that work the longitudinal bond disorder
 originated from {\em correlated} $\phi$ as in Eq. (\ref{sig.til}).
 Therefore the LRO in the present context actually corresponds to a
 logarithmically rough surface profile that is locked into a
 configuration parallel to the disordered substrate. The present
 LRO/QSRO transition corresponds there to a transition from a
 locked-in and logarithmically rough surface to a super-rough surface.

 Recently Horovitz and Golub have examined the same problem in view of
 its implications for Josephson junctions.\cite{HG97} Using a Gaussian
 variational replica approach they found a topologically similar phase
 diagram. But there are the following differences: The phase
 boundaries do not coincide exactly, since their approach does not
 take into account parameter renormalizations. In the phase where we
 find QLRO and LRO they also have QLRO and LRO. However, their phase
 with LRO is split into two sub-phases, their ``Josephson'' phase
 (where charges $\uparrow \downarrow$ are irrelevant) and
 ``Coexistence'' phase (where charges $\uparrow \downarrow$ are
 relevant). We find charges $\uparrow \downarrow$ to be relevant in
 the whole LRO phase, i.e. the whole phase with LRO has the character
 of the ``Coexistence'' phase. The boundary between their sub-phases
 coincides with the line $K^{-1}=\sigma^L$, where charges of type
 $\uparrow \downarrow$ become relevant according to the linear scaling
 analysis. However, both charges are relevant in the whole LRO phase,
 their appearance is triggered by the second-order contribution
 $y_{{\rm c} \uparrow}^2$ to the flow equation (\ref{flow.full}f) of
 $y_{{\rm c} \uparrow \downarrow}$. The most important difference is
 the region where we find QSRO and they still find QLRO that
 qualitatively influences the dependence of the critical Josephson
 current on the junction area. Again, this difference is due to the
 presence of second-order terms in the flow equations, now terms of
 order $y_{{\rm c} \uparrow \downarrow}^2$ to the flow equation
 (\ref{flow.full}f). The disagreements we find here are well-known
 from the random field $XY$ model and have to be ascribed to the
 insufficiency of the Gaussian variational ansatz to capture the
 physics represented by second-order terms to the flow equations.

\subsection{Complete picture}

 We now discuss the phase diagram in the presence of vortices {\em
 and} charges. We distinguish the cases of uniform fields and random
 fields, since their effective longitudinal bond disorder is crucially
 different.

\subsubsection{Uniform fields}

\begin{figure}
\epsfxsize=0.9 \linewidth
\epsfbox{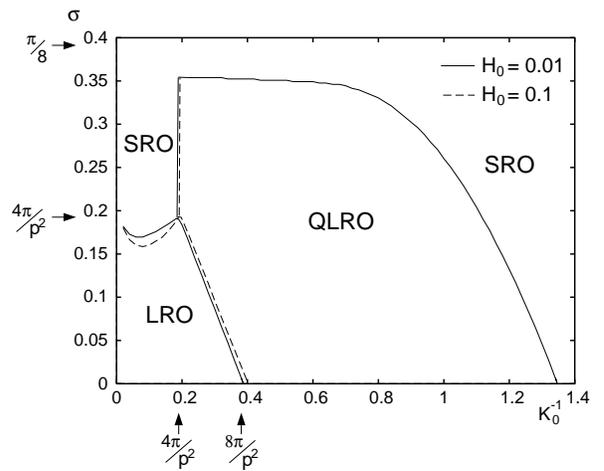}
\narrowtext
\caption{
 Phase diagram for the model with $\sigma^L=\sigma^T=\sigma$ and
 uniform fields. It is obtained from a numerical integration of the
 flow equations (\ref{flow.full}) for two different values of $H_0$
 and $\gamma=1.6$.}
\label{fig.phadi}
\end{figure}

 For definiteness we consider the case $\sigma^L = \sigma^T \equiv
 \sigma$ of random phase shifts that are uncorrelated on different
 bonds. The general case with finite $\sigma^L \neq \sigma^T $
 deviates only quantitatively. As in the absence of
 disorder,\cite{JKKN77} a phase with QLRO is stable only for
 $p>4$. For weak disorder there is a LRO phase at the low-temperature
 side of the QLRO phase.  For $4 <p< 4\sqrt 2 \approx 5.6$ LRO is
 stable up to $\sigma^L_\infty = \pi/8$, whereas for $p> 4\sqrt 2$ LRO
 is stable only up to $\sigma^L_\infty = 4\pi/p^2$. Figure
 \ref{fig.phadi} depicts the phase diagram for $p=8$, which is generic
 for all integer $p > 4\sqrt 2$. As compared to the absence of
 vortices, one main effect of vortices is the restriction of the QLRO
 phase at higher temperatures, where SRO sets in. A second important
 effect is that the phase with QSRO is turned into a phase with
 SRO. Since $\sigma$ increases logarithmically with the length scale
 in the absence of vortices, it reaches the value $\sigma=\pi/8$,
 where vortices become relevant. On small scales vortices practically
 do not contribute to a renormalization of the parameters. If we start
 from bare parameter values at a temperature slightly below the
 QSRO/QLRO transition, only charges $\uparrow \downarrow$ are
 initially relevant and increase $\sigma$ according to
 Eq. (\ref{fix.QSRO}). Then vortices become relevant only on an
 exponentially large scale\cite{CO82}
\begin{equation}
\label{L_v}
L_v \approx \exp 
\left(\frac {\pi/8 -\sigma}{(1-T/T_{\uparrow \downarrow})^2}\right) .
\end{equation}
 On this length scale SRO sets in, that is, $L_v$ represents the
 correlation length of the system. At disorder strengths $\sigma
 \lesssim \pi/8$ the phase diagram, judged by its order at largest
 scales, appears to be reentrant, i.e., a phase sequence SRO/QLRO/SRO
 can be obtained by a pure temperature change.  However, the
 difference between the low-temperature and the high-temperature part
 can be substantial in terms of the correlation length and probably
 also in terms of dynamic properties. In a remote analogy the
 high-temperature part compared to the low-temperature part of the SRO
 phase should be as different as a gas compared to a very viscous
 liquid. 

 Due to the restriction of our analysis to small fugacities of
 vortices and charges we can not rule out the possibility of an actual
 transition between two different phases at $\sigma_\infty^L \approx
 \pi/8$ for $K_\infty^{-1} \leq 4 \pi /p^2$. Since the boundary of the
 QLRO phase, which is well described in our approach, has a
 discontinuous slope at $\sigma_\infty^L \approx \pi/8$ for
 $K_\infty^{-1} \approx 4 \pi /p^2$ one might expect that this point
 is a tricritical point that should be merged by and a third
 transition line.

 For $ p < 4 \sqrt {2}$ the phase boundary LRO/SRO is determined by
 the competition of vortices and charges, which are both relevant
 already on smallest scales. The location of this transition can
 therefore not be determined. 

 The phase diagram differs from the one found in Ref. \CITE{PK91}
 crucially at low temperatures, where the SRO and QLRO phases reached
 down to the point $T=0$ and $\sigma=0$ and a reentrant phase sequence
 SRO/QLRO/LRO/QLRO/SRO was possible by changing temperature. Although
 we still find a reentrance SRO/QLRO/SRO the QLRO phase (as well as
 the LRO phase) has now a convex shape in the $(K_\infty^{1},\sigma)$
 plane.  The discrepancy can be traced back to the overestimation of
 vortex fluctuations in a vortex fugacity expansion, which is avoided
 in the present work. 

 The phase diagram Fig. \ref{fig.phadi} has been obtained by a
 numerical integration of the flow equations (\ref{flow.full}) for two
 different amplitudes $H_0=0.1$ and $0.01$ of the field and for a
 fugacity parameter $\gamma=1.6$ appropriate for the mapping of the
 original $XY$ model onto the Coulomb gases. We find only a slight
 dependence of the phase boundaries on weak fields $H_0 \ll 1$. One
 could in principle evaluate the flow equations also for larger fields
 and would probably find a direct crossover from the glassy
 low-temperature SRO region to the nonglassy high-temperature region
 as suggested by CO. However, we do not pursue this issue further
 since for $H_0 \gtrsim 1$ it is certainly no longer legitimate to
 neglect the less relevant charge classes.

\subsubsection{Random fields}
 
 The mapping discussed in Section \ref{sec.map} allows us to study
 random fields with spatially uncorrelated $\phi$ by considering
 $\tilde \sigma^L=\infty$ in the flow equations. In this limit the
 contributions proportional to $y_{{\rm c} \uparrow}$ simply disappear
 from the flow equations, in which $\tilde \sigma^L$ then no longer
 enters and the flow of $\tilde \sigma^L$ can be identified with the
 flow of $\sigma^L$ after subtracting the transformation term
 $\Delta/p^2$.

 Randomness in the field induces only a few changes in comparison to
 the case of uniform fields. The schematic phase diagram for the
 system with random fields and random phase shifts is depicted in
 Fig. \ref{fig.schem}b, which is deformed only slightly by finite
 vortex fugacities and field amplitudes.  In contrast to the case of
 uniform fields, the LRO phase disappears and the low-temperature
 QLRO/SRO boundary is located at $K_\infty^{-1}=4\pi/p^2$ down to
 $\sigma=0$.  Thus a phase with QLRO now exists for $p >2 \sqrt
 2$.\cite{CO82} The spin correlation function has a finite exponent
 (\ref{eta.inf}) (involving the original $\sigma^L$ and not $\tilde
 \sigma^L$ since we are interested in the correlation of $\theta$ and
 not of $\tilde \theta$).

 For temperatures $T \lesssim T_{\uparrow \downarrow}$ the
 Cardy-Ostlund charges $\uparrow \downarrow$ become relevant and still
 generate transverse bond disorder according to Eq. (\ref{fix.QSRO})
 (which also describes the further decrease of longitudinal bond
 disorder). Therefore vortices become relevant again on scales
 (\ref{L_v}) for $T \lesssim T_{\uparrow \downarrow}$.

\section{Discussion}

 To summarize, we have performed a renormalization group analysis of
 the model (\ref{H}) in the Villain approximation that allowed us to
 represent the model by two coupled Coulomb gases. The breakdown of
 the fugacity expansion for the vortex gas\cite{seK93,MW97} was
 treated nonperturbatively (the flow equations contain the effective
 vortex fugacities $y_{\rm v}$ that are nonperturbative functions of
 the original vortex fugacity $\sim
 e^{-1/K}$).\cite{NSKL95,CF95,S97,lhT96} We have shown that an
 analogous breakdown does {\em not} occur for an expansion in the
 charge fugacities and have identified two most relevant classes of
 vector charges out of an infinite set. The renormalization group flow
 equations for the interacting Coulomb gases have been established and
 evaluated for the limit where both gases are dilute, i.e., for low
 temperatures and weak fields.

 From these flow equations we have determined the structure of the
 phase diagrams, in particular the domains of stability of the phase
 with QLRO. Due to the breakdown of the fugacity expansion for the
 vortices this phase was found to be more stable against disorder than
 found previously.\cite{CO82,PK91} The critical exponents
 $\eta_\infty$, $\overline{\eta}_\infty$, and $\delta$ have been given
 in Eqs. (\ref{eta.inf}), (\ref{overeta.inf}), and (\ref{delta.inf}).

 In a strict sense our approach is valid only for the characterization
 of the QLRO phase, where vortices and fields are asymptotically
 irrelevant. Nevertheless we have attempted to complete the picture of
 the phase diagram by examining how vortices and fields become
 relevant, which was taken as indication for the LRO, QSRO, or SRO
 nature of the neighboring phases. Thereby we were also able to locate
 qualitatively the LRO/QLRO transition in the case of uniform
 fields. This transition was found to merge one corner of the phase
 with QLRO.  Its location is in qualitative agreement with that found
 in a recent Gaussian variational approach,\cite{HG97} where vortices
 had not been considered. On the high-temperature side of the QLRO
 phase vortices become relevant on relatively short length scales and
 immediately lead to an increase of the effective temperature and a
 suppression of the charge fugacities.  On the low-temperature side of
 the QLRO phase charges become relevant first. For weak longitudinal
 disorder ($\tilde \sigma^L < 4 \pi/p^2$) the flow converges to a LRO
 fixed point, otherwise the flow tends on intermediate scales towards
 the QSRO fixed point before it eventually crosses over to the SRO
 fixed point. Since the crossover from QSRO to SRO occurs at finite
 values of the charge fugacities, this behavior might be modified by
 higher orders in the charge fugacities, which we are not able to
 incorporate systematically at present. This difficulty excludes also
 the calculation of phase diagrams for uniform fields with $p<4$ and
 random fields with $p < 2 \sqrt 2$ in the present framework.

 CO have suggested the possibility of a first-order transition between
 the glassy low-temperature SRO and the based on the nonglassy
 high-temperature SRO. To our understanding this scenario was based on
 low-temperature properties of the vortex fugacity flow equation that
 are absent in the present analysis, where the vortex fugacity
 expansion has been avoided.

 We speculate that there could be a transition from a low-temperature
 SRO phase to a high-temperature SRO phase that merges the upper left
 corner of the QLRO phase. It could be possible that this transition
 line persists even at very strong disorder and ends up at
 $\sigma=\infty$ and $T=0$. This point corresponds to the so-called
 gauge glass model that represents a zero-temperature critical
 point.\cite{gg} However, for such a transition line we have no
 evidence other than the fact that corners usually occur only at
 a three-phase coexistence and that the gauge glass fixed point should
 have a continuation for weaker disorder (finite $\sigma$). It would
 be interesting to examine this scenario further, which probably can
 be achieved only by numerical methods.

 The studied model has physical applications in various fields. The
 most direct application is to $XY$ magnets, where random phase disorder
 can be provided by nonmagnetic impurities through the
 Dzyaloshinskii-Moriya interaction and local crystal fields break the
 spin rotation symmetry.\cite{RSN83}

 In the case when vortices are excluded, the model also describes
 crystal surfaces grown on a disordered substrate.\cite{TV90,S95} Such
 disorder can also be related to dislocations that leave the crystal
 at the surface.\cite{BA97} The LRO/QSRO transition described above
 corresponds to a disorder driven transition from a rough crystal
 surface to a super-rough phase. In the context of planar Josephson
 junctions\cite{HG97} LRO corresponds to a finite critical current
 density even for arbitrarily large contacts and QSRO corresponds to a
 critical current density that decays faster than algebraically with
 the contact area.

 The phase diagrams obtained here should also be qualitatively similar
 to those of crystalline films on periodic substrates with
 disorder\cite{pS97,CD97a} and of vortex lattices.\cite{CD97} The
 transitions occur between phases that differ in the degree of
 structural order. However, these systems are more complicate to
 analyze in a similar framework since they have a two-component
 displacement field instead of the one-component phase of model
 (\ref{H}).  In addition, the periodicity of the lattice implies that
 both the particle interactions and the pinning energy have the same
 wavelength, i.e., $p=1$. Uniform and disordered fields then
 correspond to pinning that is commensurate or incommensurate with the
 crystal. Since $p=1$ it appears likely that there is no phase where
 the effects of pinning and the proliferation of dislocations are
 irrelevant on large scales and that the strongly relevant fields and
 dislocations cannot be well described by a direct generalization of
 the present framework.

\section*{Acknowledgments}

We are grateful to J. Kierfeld and T. Nattermann for stimulating
discussions and a critical reading of the manuscript.

S.S. acknowledges support from the Deutsche Forschungsgemeinschaft
under Project No. SFB341 and Grant No. SCHE/513/2-1 and from the
NSF-Office of Science and Technology Centers under contract No.
DMR91-20000 Science and Technology Center for Superconductivity.

\appendix

\section{The Coulomb gas mapping}
\label{app.cg}

 In this appendix some intermediate steps in the derivation of the
 Coulomb gas model (\ref{H.rs}) are given.  The starting point is the
 effective Hamiltonian (\ref{H.vil}). The first step is to replace the
 scalar field $m$ by the charge field $n$ with the use of the
 Poisson summation formula:
\begin{eqnarray}
\label{H.1}
{\cal H}&=& \sum_{\bf r} \Big\{ \frac {K}2
({\bbox \nabla} \theta_{\bf r} - {\bf A}_{\bf r}+2 \pi {\bf M}_{\bf r})^2
\nonumber\\
&& + \frac 1 {2 H} n_{\bf r}^2 +i p n_{\bf r} \theta_{\bf r} \Big\}.
\end{eqnarray}
 In the corresponding partition function all $\theta \in [0,2 \pi]$ are
 integrated over and all ${\bf M} \in {\mathbb Z}^2$ and $n \in
 {\mathbb Z}$ are summed over. The integrations can be extended to
 $\theta \in {\mathbb R}$ if the summation over the ${\bf M}$ is
 restricted to a summation over a subset $\hat {\bf M}$. The selection
 of the subset fixes a gauge for ${\bf M}$, which is the vector
 potential of the vortex density, $N={\bbox \nabla} \times \hat{\bf
 M}$. The summation over the subset of $\hat {\bf M}$ then corresponds
 to a summation over the vortex density.  

 The integration over the angles is most conveniently performed in
 Fourier space and results in the Hamiltonian (\ref{H.rs}). In the
 following we examine closer the vortex-charge interaction that reads
 on at an intermediate level
\begin{equation}
\label{H.2}
{\cal H}_{\rm vc}=2 \pi i p \int_{\bf k} \frac 1{k^2}
n_{- \bf k} i {\bf k} \cdot \hat {\bf M }_{\bf k}.
\end{equation}
 Note that it is not possible to choose a Coulomb gauge with $ {\bf k}
 \cdot \hat {\bf M}_{\bf k}=0$ due to the integer nature of ${\bf
 M}_{\bf r}$. The statistical weights $\exp(-{\cal H})$ are invariant
 under a change of gauge $\hat {\bf M} \to \hat {\bf M} + \chi$ due to
 the integer nature of the charge field $n_{\bf r}$ and the gauge
 field $\chi_{\bf r}$. To transform expression (\ref{H.2}) back into
 real space, let us look for simplicity at the ground state of
 Hamiltonian (\ref{H.1}) in the absence of fields and disorder,
\begin{equation}
\theta^{(0)}_{\bf k}=\frac {2 \pi i} {k^2} \hat {\bf M}_{\bf k}.
\end{equation}
 But this is just the phase field generated by vortices, which can be
 written in real space (neglecting discretization effects)
\begin{equation}
\theta^{(0)}_{\bf r}=\sum_{\bf R} \omega({\bf r}- {\bf R}) N_{\bf R},
\end{equation}
 where $\omega({\bf r}- {\bf R})$ measures the angle enclosed between
 the vector $({\bf r}- {\bf R})$ and an arbitrary reference direction
 in the $xy$ plane. In direct analogy the coupling (\ref{H.2}) becomes
 (\ref{H.vc}) in real space. $\theta^{(0)}$ and ${\cal H}_{\rm vc}$ do
 not depend on the choice of this reference direction for vortex
 configurations satisfying the neutrality (\ref{neutr}). Only these
 configurations have finite energy and contribute to the statistics.

\section{Screening of charges by vortices}
\label{app.screen}

 Here we show how the angular interaction between vortices and charges
 induces a screening (\ref{Kc.screen}) of the interaction among
 charges when vortices are integrated out from the partition sum.
 Following Kosterlitz\cite{jmK74} and Young\cite{apY79} we integrate
 out every pair of vector vortices with radius less than $1+dl$.  The
 contribution of a vortex pair with ${\bf N}_1+{\bf N}_2 =0$ to the
 Hamiltonian ${\cal H}_{\rm vc}$ is
\begin{equation}
  {\cal H}_{\rm vc} = ip \sum_{\bf r} {\bf N}_1 \cdot {\bf n}_{\bf r} [
   \omega({\bf R_1}-{\bf r})-\omega({\bf R_2}-{\bf r})].
\end{equation}
 Since we assume a diluted Coulomb gas, so that typically $\left|{\bf
 R}_1-{\bf r}\right|\gg \left|{\bf R}_1-{\bf R}_2\right|$, one can
 expand the parts of the partition sum including this pair:
\begin{eqnarray}
  \lefteqn{\exp\left(-{\cal H}_{\rm vc}\right) \approx  1-\frac{p^2}{2}
  \sum_{{\bf r},{\bf r}'}  {\bf n}_{\bf r} \cdot {\bf n}_{{\bf r}'}}
\nonumber\\
  & &\times [\omega({\bf R}_1-{\bf r})-\omega({\bf R}_2-{\bf r})]
  [\omega({\bf R}_1-{\bf r}')-\omega({\bf R}_2-{\bf r}')].
\end{eqnarray}
 In contrast to the screening effects within the vortices, the first
 nontrivial term in this expansion, which is even in the charge
 density and produces nonvanishing screening, is of second order in
 the Hamiltonian. In the dilute limit we can expand
\begin{equation}
  \omega\left({\bf R}_1-{\bf r}\right)-\omega\left({\bf R}_2-{\bf r}\right)
  \approx
  \frac{({\bf R}_1-{\bf r})\times({\bf R}_2-{\bf r})}{\left|{\bf R}_1-
  {\bf r}\right|\left|{\bf R}_2-{\bf r}\right|}.
\end{equation}
 Then it is straightforward to perform the integration over the
 annulus $1\leq |{\bf R}_1 - {\bf R}_2| < 1+dl$. After that step one
 can translate the vortex pair over the whole space excluding the
 spheres occupied by the other vector pairs and obtains a screening of
 charges by vortices very similar to the screening of vortices by
 vortices.\cite{mL97}

\section{Dictionary}

For a convenient comparison of the present work with CO we include the
following substitution list for symbols (restricted to the replica
limit $n \to 0$).
\begin{equation}
\label{dict.CO}
\begin{array}{rcl}
\mbox{CO symbol} &\to& \mbox{symbol of present work}:
\\
\tilde K &\to& K,
\\
K &\to& K-\sigma^T K^2,
\\
\tilde h_p &\to& H_0,
\\
\vec n &\to& {\bf n}/\sqrt 2,
\\
y_p &\to& e^{1/H} \approx \frac {H_0^2}4 \approx  
y_{{\rm c} \uparrow \downarrow},
\\
\tilde y &\to& \sqrt{\pi} p y_{{\rm c} \uparrow \downarrow},
\\
y^2 &\to& 4 \pi^3 y_{\rm v}^2.
\end{array}
\end{equation}

\newpage

\end{multicols}


\begin{references}
\bibitem{note.bonds}
For an explicit discussion of the generation of random bond disorder
by random fields and its effect on the spin correlation see also Refs. 
\CITE{GH82,VF84,GS85}.
\bibitem{note.comp.JKKN77}
$y_{\rm v}$ becomes the usual vortex fugacity and $y_{{\rm c}\uparrow
\downarrow}$ represents negligible higher orders.  The agreement with
the corrected flow equations Eq. (5.17) of Ref. \CITE{JKKN77} is given
if one omits the exponential factors, which we understand as attempt
to integrate the initial values for the fugacities.
\bibitem{note.comp.CO82}
The agreement is given if one replaces the factor multiplies the
factor 4 in Eq. (3.7) of Ref. \CITE{CO82} by an additional factor
$\pi$ [see dictionary (\ref{dict.CO})].
\bibitem{note.comp.HG97}
However, there are nontrivial discrepancies in the second-order
contributions to the flow equations (22) of Ref. \CITE{HG97}.
\end{references}
\end{document}